\documentclass{egpubl}
\usepackage{eurovis2024}

 \STAREurovis   % for EuroVis additional material 
\CGFccbyncnd

\usepackage[T1]{fontenc}
\usepackage{dfadobe}
\usepackage{cite}  % comment out for biblatex with backend=biber
\BibtexOrBiblatex
\electronicVersion
\PrintedOrElectronic
\usepackage[pdftex]{graphicx}
\usepackage{egweblnk}
\usepackage{enumitem}
\PassOptionsToPackage{svgnames}{xcolor}
\usepackage{svg}
\usepackage{wrapfig}
\usepackage{scalerel}
\usepackage{soul}
\usepackage{xcolor}
\usepackage{xspace}

\DeclareRobustCommand{\ctext}[2]{{\sethlcolor{#1}\hl{#2}}}

\newcommand{\highlight}[1]{{{#1}}}

\definecolor{papercolor}{HTML}{E6E6E6}
\definecolor{whycolor}{HTML}{D7EDEB}
\definecolor{whatcolor}{HTML}{FAF0DA}
\definecolor{howcolor}{HTML}{FADDD6}
\newcommand{\paper}[1]{\ctext{papercolor}{\texttt{#1}}}
\newcommand{\why}[1]{\ctext{whycolor}{\texttt{#1}}}
\newcommand{\what}[1]{\ctext{whatcolor}{\texttt{#1}}}
\newcommand{\how}[1]{\ctext{howcolor}{\texttt{#1}}}

\newcommand{\location}{\scalerel*{\includegraphics{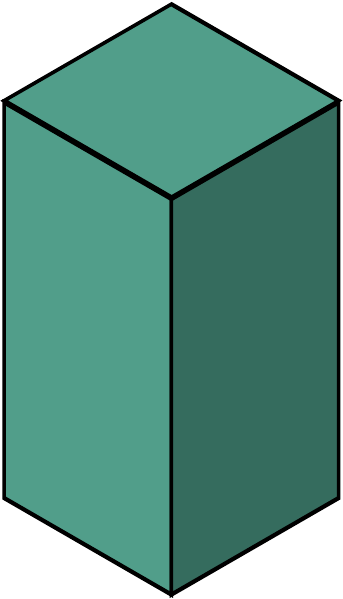}}{B}\xspace}
\newcommand{\target}{\scalerel*{\includegraphics{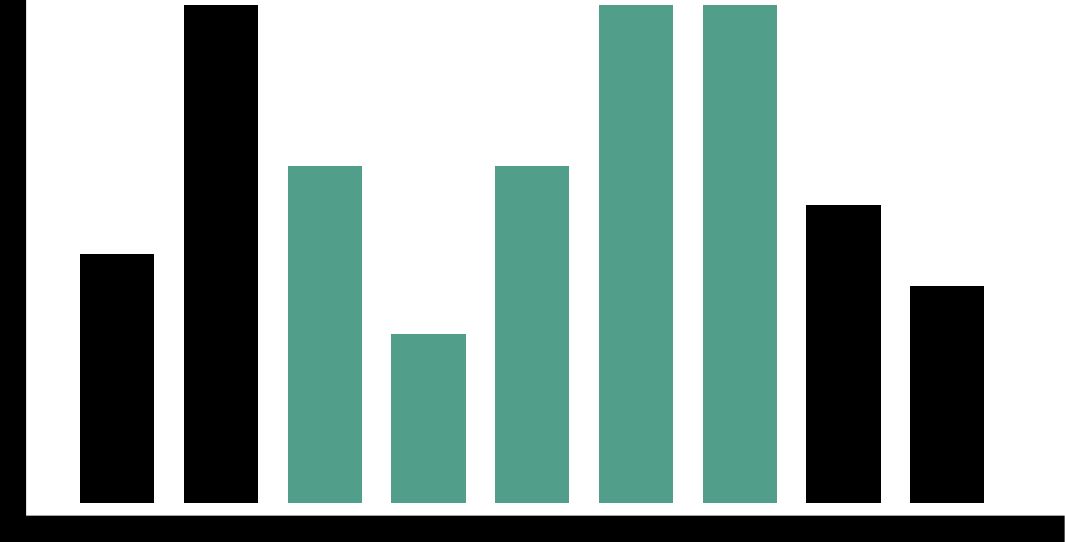}}{B}\xspace}
\newcommand{\nolocation}{\scalerel*{\includegraphics{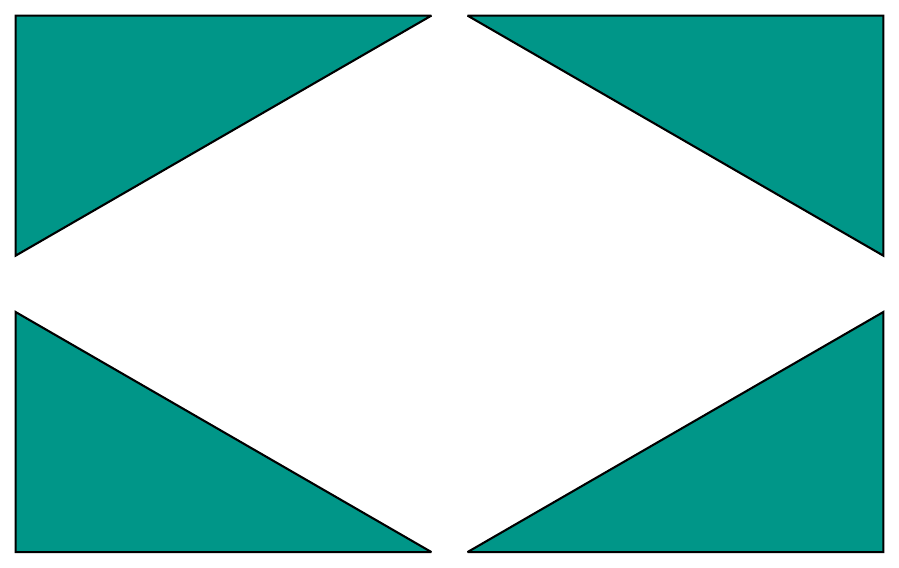}}{B}\xspace}
\newcommand{\notarget}{\scalerel*{\includegraphics{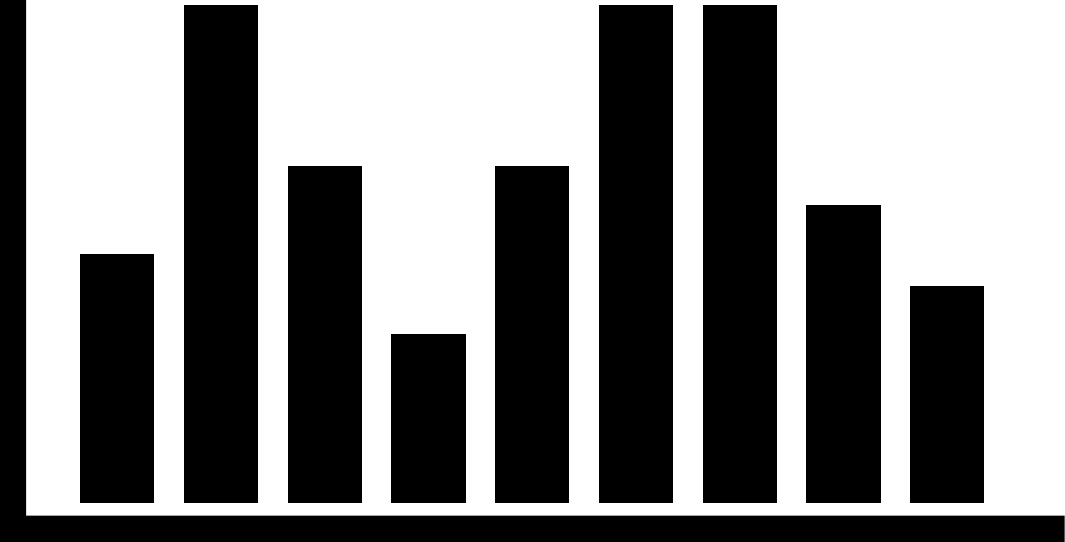}}{B}\xspace}

\title[The State of the Art in Visual Analytics for 3D Urban Data]%
      {The State of the Art in Visual Analytics for 3D Urban Data}

\author[Miranda et al.]
{\parbox{\textwidth}{\centering Fabio Miranda$^{1}$\orcid{0000-0001-8612-5805}, Thomas Ortner\orcid{0000-0002-9373-6409}, Gustavo Moreira$^{1}$\orcid{0000-0002-4762-7703}, Maryam Hosseini$^{2}$\orcid{0000-0001-8329-4638}, Milena Vuckovic$^{3}$\orcid{0000-0002-5825-8237},\\ Filip Biljecki$^{4}$\orcid{0000-0002-6229-7749}, Claudio T. Silva$^{5}$\orcid{0000-0003-2452-2295}, Marcos Lage$^{6}$\orcid{0000-0003-3868-8886}, Nivan Ferreira$^{7}$\orcid{0000-0001-6631-4609}
        }
        \\
{\parbox{\textwidth}{\centering $^1$Department of Computer Science, University of Illinois Chicago, USA\\
         $^2$City Form Lab, Massachusetts Institute of Technology, USA\\
         $^3$VRVis Zentrum f\"ur Virtual Reality und Visualisierung Forschungs-GmbH, Austria\\
         $^4$Department of Architecture and Department of Real Estate, National University of Singapore, Singapore\\
         $^5$Tandon School of  Engineering and Center for Data Science, New York University, USA\\
         $^6$Institudo de Computa\c{c}\~{a}o, Universidade Federal Fluminense, Brazil\\
         $^7$Centro de Inform\'{a}tica, Universidade Federal de Pernambuco, Brazil
       }
}
}

\begin{document}

% uncomment for using teaser
% \teaser{
%  \includegraphics[width=\linewidth]{eg_new}
%  \centering
%   \caption{New EG Logo}
% \label{fig:teaser}
%}

\maketitle
%-------------------------------------------------------------------------
\begin{abstract}
Urbanization has amplified the importance of three-dimensional structures in urban environments for a wide range of phenomena that are of significant interest to diverse stakeholders. 
With the growing availability of 3D urban data, numerous studies have focused on developing visual analysis techniques tailored to the unique characteristics of urban environments. However, incorporating the third dimension into visual analytics introduces additional challenges in designing effective visual tools to tackle urban data's diverse complexities.
In this paper, we present a survey on visual analytics of 3D urban data. Our work characterizes published works along three main dimensions (\emph{why}, \emph{what}, and \emph{how}), considering use cases, analysis tasks, data, visualizations, and interactions.
We provide a fine-grained categorization of published works from visualization journals and conferences, as well as from a myriad of urban domains, including urban planning, architecture, and engineering.
By incorporating perspectives from both urban and visualization experts, we identify literature gaps, motivate visualization researchers to understand challenges and opportunities, and indicate future research directions.


\end{abstract} 
%-------------------------------------------------------------------------

\setlength{\intextsep}{0pt}%
\setlength{\columnsep}{5pt}%

\section{Introduction}

Over the past few decades, regions around the world have undergone a rapid process of urbanization, resulting in the need for cities to \emph{densify} to meet rising housing demands.
%
% Cognizant of this trend and the fact that cities are characterized by their verticality, a growing number of domains have gone beyond the usual flatland that defines a spatial region.
\highlight{In response to this trend and recognizing the multifaceted and dynamic nature of urban environments, there has been a movement from the traditional use of 2D data and representations towards approaches that better acknowledge the three-dimensional aspects of cities.}
Many phenomena of interest to a variety of stakeholders, such as civil engineers, urban planners, architects, and climate scientists, are inherently three-dimensional, requiring reasoning over the 3D structure of urban environments.
These domains have progressively adopted the third dimension in many of their analytical tasks to study and tackle urban problems.
These tasks often rely on data that is intrinsic to the \emph{physical} aspect of cities.
The transition to more sustainable environments, energy sources, and technologies has underscored the importance of leveraging this 3D structure in its entirety.
In another front, the advancements of technologies such as virtual reality (VR) and augmented reality (AR) have created opportunities for new experiences in exploring 3D urban environments~\cite{chen_exploring_2017,chen_immersive_2017,chen_urbanrama_2022}. 

However, the inclusion of this additional dimension increases the difficulty in addressing the various challenges involved in designing effective GIS and visual analytics tools. These tools require visual strategies to support analysis of the data referent to the city's geometry (i.e., 3D urban data), navigation to learn the structure of the environment and integration of the information from different points of view, while avoiding common problems such as occlusion and higher cognitive load associated with frequent viewpoint changes and lack of sensorial stimuli~\cite{elmqvist_tour_2007,burigat_navigation_2007}. 
\emph{
Tackling these challenges can be fundamental to uncovering features valuable for decision-making and problem-solving in several domains.}

But how do domain experts analyze 3D urban data? What are the visual analytics techniques and tools being used by practitioners and experts? To answer such questions and evaluate the state of the art in 3D urban data analytics, in this survey, we have reviewed over 20 journals and conferences from 2008 to 2023, including visualization and cross-cutting multi-disciplinary ones.
Our focus is on the analysis tasks, visual analytics tools, applications, and visualization techniques that propose to tackle problems and enable tasks involving 3D urban data. 
Our survey bridges the knowledge gap between different communities, including visualization, urban planning, architecture, and engineering, and helps identify research challenges that can benefit from these multiple perspectives.
Given the diversity of use cases, data, and tasks and the lack of empirical studies on the topic, our primary goal with this survey is to inform the visualization community about challenges and opportunities.
We believe that this can foster advancements in both theoretical and applied research in this field and generate a set of well-grounded and concrete recommendations in the future; we hope to inform the visualization community of the challenges and opportunities in 3D urban visual analytics, as well as common needs that urban domain experts have.
Our contributions are summarized as follows:

\begin{itemize}
\item We first establish a common characterization that allows us to organize contributions from a multitude of domains, including visualization, architecture, engineering, and urban planning.
\item We introduce a comprehensive survey on 3D urban visual analytics. Inspired by other surveys \cite{dani_ten_2019, pandey_state_2021, clarinval_intra_2021}, we follow a human-centered, interrogative method in which we classify each evaluated work with respect to Munzner's~\cite{munzner_visualization_2015} analytical framework of \emph{\textbf{Why} is 3D urban data being analyzed}, \emph{\textbf{What} data is being analyzed}, and \emph{\textbf{How} it is being analyzed and visualized}.
%
% This enables a fluid discussion on the contribution of existing works.
% e establish a common classification process to organize contributions from a multitude of domains, including visualization, GIS, architecture, and urban planning. Research efforts in these communities are often not aware of one another, and through our classification 
% Urban 3D analytics is an important tool in multiple domains. For this reason, it is not restricted to visualization venues. Rather, it also includes research efforts from, for example, GIS and urban planning communities. As discussed later, research efforts in these communities often are not aware of one another, and therefore there is a need to connect the solutions produced.
\item We report a series of research directions and open problems in 3D urban visual analytics. These include visual metaphors, navigation \& occlusion, data and systems integration, evaluation of visual designs, and guided explorations.
\end{itemize}
%

%
% \nivan{review this}
%
Our survey is organized as follows: Section~\ref{sec:motivation} presents a background on 3D urban analytics and data. In Section~\ref{sec:related}, we review related surveys. In Section~\ref{sec:methodology}, we introduce our survey scope and methodology. Section~\ref{sec:overview} presents an overview of the survey, and Section~\ref{sec:paper_types} presents and discusses the types of papers covered in our survey.
In Sections~\ref{sec:why} and \ref{sec:what}, we group and discuss papers according to primary (i.e., \textbf{Why}) and data (i.e., \textbf{What}) dimensions. In Section~\ref{sec:tasks}, we list common tasks across use cases and, in Section~\ref{sec:how}, we discuss \textbf{how} visualization and interaction techniques facilitate these tasks.
%
% In Section~\ref{sec:takeaways}, we provide a series of takeaways for visualization researchers and in Section~\ref{sec:challenges}, we identify challenges and opportunities for visual analytics of 3D urban data. 
%
In Section~\ref{sec:new_challenges}, we discuss the main observations from our survey and we identify challenges and opportunities for visual analytics of 3D urban data. 
Section~\ref{sec:conclusion} concludes our survey.
%
% \nivan{make sure we added the new references here}
The full list of reviewed papers can be accessed at \url{https://urbantk.org/survey-3d}.

\begin{figure*}[ht!]
\begin{center}
\includegraphics[width=1\linewidth]{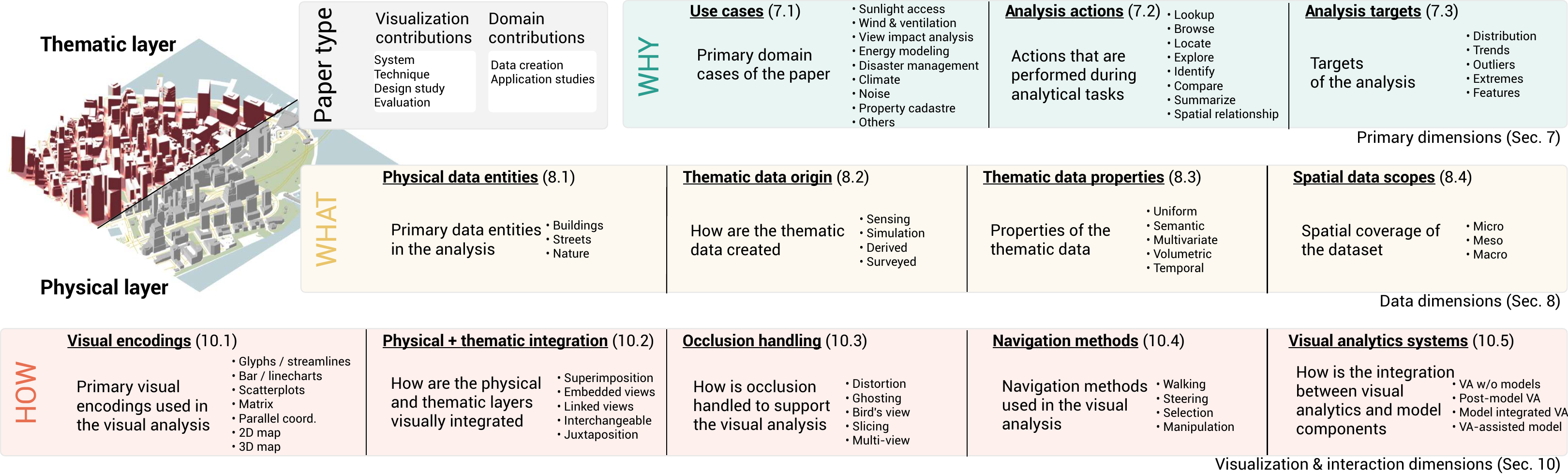}
\end{center}
% \vspace{-0.5cm}
\caption{Overview of the survey, centered around paper type and three main dimensions asking \why{Why}, \what{What}, and \how{How} visualization supports 3D urban data analytics. Each dimension corresponds to a section of the survey, and major subsections highlight the specific tags considered in each category. We also characterize papers according to their primary display modality and evaluation method. Top left: an illustration of the thematic layer (3D urban data with sunlight access information) and the physical layer (physical form of Downtown New York City).}
% \vspace{-0.5cm}
\label{fig:overview}
\end{figure*}

%Zoning regulations : their impacts and alternative scenario planning, community engagement and outreach 

\section{Background}
\label{sec:motivation}

% \nivan{maybe we should mix this with the intro and not have this as a new section?}

%\fabio{somewhere here it needs to explain the different layers}

% \nivan{We talk about this already in the intro, but I wonder if we could include something abour the challenges of visualizing 3d urban data here?}

\noindent \textbf{3D urban analytics.}
Cities are three-dimensional in nature and hence, should be described and analyzed in three dimensions, considering their structures and urban forms.
For instance, the presence of high-rise buildings affects wireless connectivity, while the height of buildings, the density of roads, and the amount of green spaces influence environmental conditions like air quality and temperature~\cite{hankey_urban_2017,cabral_designing_2023}.
\highlight{However, in the absence of appropriate methods and techniques to analyze 3D data,} urban morphology analysis has adopted various 2D measures to describe the physical form of cities~\cite{2022ceusgbmi}. These measures often fail to draw a vivid picture of the reality of urban life, leading to simplified models, calculations, and, ultimately, policies that lack the precision needed to translate findings into actionable insights~\cite{shi_semantic_2021}. 

The physical form of cities in many parts of the world is largely influenced by a set of regulations known as zoning laws, which dictate, among others, the size, shape, and bulk of the buildings. Although inherently a three-dimensional concept, zoning codes are conventionally enacted without leveraging 3D information. This makes it very difficult to assess the large-scale impact of new regulations on varying aspects of urban life ranging from wind patterns~\cite{lindberg_towards_2005, deininger_continuous_2020, brasebin_3d_2018}, air quality~\cite{zhang_exploring_2022, casazza_3d_2019}, and heat emission~\cite{dissegna_3-d_2019, yuan_mitigating_2020} to the flow of people, goods, and services~\cite{lv_smart_2022, zhang_walking_2022}. 3D models provide a more realistic view of these impacts, empowering local communities to understand better the implications of zoning codes on their everyday life~\cite{brasebin_3d_2018}.  

The use of 3D analysis in urban planning and policy also extends to other areas, such as urban ecology~\cite{tian_effect_2019}. 
For example, the Urban Heat Island (UHI) effect, which is characterized by a sharp rise in urban temperature, poses a significant threat to public health, the ecological environment, and urban livability~\cite{estoque_effects_2017, huang_investigating_2019}. 
3D models can help mitigate the UHI effect by providing a deeper understanding of factors such as the sky view, building bulk, and horizontal reflectance~\cite{chun_daytime_2017, gal_computing_2009, berger_spatio-temporal_2017, gautier_visualizing_2020, rink_environmental_2022}. Additionally, 3D analysis is crucial in the realm of urban disaster management, enabling the accurate identification of catchment areas in events such as flooding~\cite{wang_flood_2019}, earthquakes~\cite{redweik_3d_2017}, or even the outbreaks of infectious disease~\cite{megahed_antivirus_2020, eremchenko_infectious_2020, salaheldin_microclimatic_2021, xu_modelling_2023}.

Urban digital twins, which have gained increasing attention in recent years, provide a virtual representation of cities as complex systems~\cite{arcaute_future_2021,charitonidou_urban_2022,2023autcondtchallenges}. These digital replicas usually employ 3D visualization to bring together stakeholders and various sources of heterogeneous data, models, and algorithms describing various parts of urban systems such as urban ecology~\cite{schrotter_digital_2020}, transportation~\cite{jiang_digital_2022}, and economic and social functions~\cite{batty_digital_2018, dembski_urban_2020, arcaute_future_2021}. They enable 3D spatio-temporal simulations and impact assessment analysis. Urban digital twins can inform decision-makers about the multi-scale impacts of policies as well as empower participatory design and community involvement in urban planning~\cite{wan_developing_2019, dembski_digital_2019}.
While previous works have extensively leveraged 2D screens, VR and AR technologies also have the potential to offer powerful data analysis capabilities while providing an immersive experience~\cite{chen_exploring_2017}.
These systems open avenues for novel interactions to enhance visual data analysis in 3D urban environments, by providing new strategies for city navigation~\cite{chen_urbanrama_2022} and to deal with building occlusion~\cite{chen_immersive_2017}.

% \begin{figure}[b!]
% \begin{center}
%     \includegraphics[width=\linewidth]{figs/cycle.pdf}
% \end{center}
% % \vspace{-0.5cm}
% \caption{Stages of the urban data life cycle. \thomas{I am still skeptical what benefit this figure gives us. It might even be misleading about what parts of this pipeline we cover. One of the reviewers also pointed out it is rather generic}}
% \label{fig:cycle}
% \end{figure}

\noindent \textbf{3D urban data.}
%
% \nivan{R1 commented that this is too general and maybe should be removed`} \thomas{I think it would not hurt just to remove it unless we strongly tie later sections to it} \fabio{Yes, I think we should remove it -- just realized it these stages never come up again anywhere else in the paper}
Urban visual analytics is composed of a set of stages to process, analyze, and visualize data. In the context of 3D urban data, the first stage involves the collection of 3D data describing the physical layers.
\highlight{This data can originate from a variety of sources, including authoritative city agencies and crowdsourced initiatives like OpenStreetMap~\cite{haklay_open_2008}. The methods employed to acquire this data encompass a range of techniques, such as sensing with LiDAR or mobile mapping.}
\highlight{In this survey, we define 3D urban data as the information inherently associated with the three-dimensional structure of urban environments.}
Depending on the use case, such data can also contain attributes describing specific properties of the physical form of the city.
The data is then transformed to satisfy analysis, modeling, or visualization requirements. For example, transforming OpenStreetMap data into polygon meshes and modifying mesh resolutions.
Transformed data is optionally stored in general (e.g., MySQL) or 3D-specific (e.g., 3DCityDB) databases to facilitate querying. 
Finally, the physical layer data is leveraged for analysis and modeling to create new thematic layers, which are then visualized.

Designing visual analytics tools for experts considering 3D urban environments and digital twins poses several challenges along this pipeline. Data acquisition and transformation require the integration and management of disparate data formats (e.g., building footprints, tabular data, heightfields, street networks).
Modeling certain phenomena oftentimes requires complex simulations, the involvement of multiple stakeholders, and deep domain expertise.
Finally, the analysis and visualization of data in 3D urban scenarios require strategies to map thematic data to the urban environment. This is made more challenging if the data is temporal and/or multivariate.  
On top of this, tools should also consider strategies to facilitate navigation in the 3D environment and minimize occlusion, both across buildings and within the same building.

\highlight{While it is also crucial to acknowledge that data visualization in general remains separate from the dimensionality of the data, it is equally important to recognize that visualizing 3D data within a 2D design space can lead to substantial information loss due to its inherent constraints. This stems from overplotting, visual clutter and occlusion. This occurs primarily because numerous data points may be superimposed over each other, making it challenging for a human analyst to discern individual data points and infer potential spatial relationships. Particularly in contexts where depth, spatial relationships and immersion are critical to understanding complex 3D geospatial data and its features, 3D visualizations offer distinct advantages over traditional 2D counterparts.}

\noindent \textbf{Outlook.}
In this survey, we organize the targets of analysis into two \emph{layers}. 
The physical layer represents the physical form of the city and its geometries. 
The other one called the thematic layer, stores the 3D urban data that is output by simulations, machine learning models, sensing initiatives, or surveys. 
Figure~\ref{fig:overview} (top left) illustrates the thematic and physical layers.
At the center of our work is the survey of contributions that leveraged these layers in their studies.

%digital twins of the cities 
% As discussed, 3D enables more realistic simulations of urban life. Urban digital twins, which received a lot of attention in recent years, combine various sources of heterogeneous data and bring them to life through 3D visualization to create digital replicas of cities~\cite{dembski2020urban}. They offer a virtual representation of cities as complex systems, incorporating a variety of different features and functions including the ones described earlier~\cite{}. 
% % Digital twins contains models, algorithms,  
% To be effective, they require rigorous modeling and analysis for 3D transformation of geospatial data, accounting for surface terrain~\cite{}, 
% economics and social functions~\cite{batty_digital_2018, arcaute_future_2021}. 
% Apart from informing decision-makers about the multi-scale impacts of policies on different aspects of urban systems, digital twins can empower participatory design and more effective community involvement in urban planning~\cite{dembski_digital_2019}. 

%3D spatio-temporal simulations
%

% detailed, complex, and The digital twins models unleash powerful 

\section{Related surveys}
\label{sec:related}

Despite the growing body of literature on 3D urban visual analytics, no comprehensive survey has been published on it.  
% in the space of 3D urban visual analytics is growing, 
% there has not been a comprehensive survey on this topic. 
%
However, there are surveys on related topics. For example, Kraus et al. reviewed immersive analytics and the use of 3D techniques for visualization in general, not focusing on urban data~\cite{kraus_immersive_2022}. Moreover, the study excluded papers where visualizations are inspected on 2D screens (therefore not covering a large body of work that relies on this modality). 
Chen~et~al., on the other hand, reviewed the contribution of visualization for urban analytics to derive a design space for immersive urban analytics systems~\cite{chen_exploring_2017}. Nevertheless, the study does not include important dimensions explored in our survey (e.g., analytical tasks, occlusion handling techniques) and also excludes relevant work done in other research communities.
Doraiswamy et al. presented a high-level overview of the challenges of urban data, focusing on 2D-based contributions, with only a brief discussion of the 3D opportunities~\cite{doraiswamy_spatio_2018}.
Other urban analytics surveys, such as Deng et al.~\cite{deng_survey_2023}, Feng et al.~\cite{feng_survey_2022}, and Zheng et al.~\cite{zheng_visual_2016}, did not cover 3D urban data and, more importantly, visualizations and tasks specifically aimed at that type of data.

In summary, this paper aims to fill the existing gaps by reviewing the literature on 3D urban data, analytics tasks, and use cases, surveying current visualization techniques in that space, and contributing a classification to allow connections and collaborations between urban and visualization experts.

\begin{table}[t]
\caption{The main venues reviewed in this survey (horizontal line separates the visualization-related ones and domain application venues, with journals shown first). Papers that appeared between 2008 and 2023 were reviewed and the ones related to 3D urban analytics were included in our analysis. The citations in those papers were also used as sources of relevant prior research.}
% \vspace{-0.25cm}
\centering
\small
\begin{tabular}{l|l}
\hline
 CFG    &  Computer Graphics Forum\\
CG   & Computers \& Graphics \\
  CGA  & IEEE Computer Graphics and Applications\\
  IV   & Information Visualization\\  
TOG  & ACM Trans. on Graphics \\
TVCG & IEEE Trans. on Visualization and Computer Graphics \\
TVCJ & The Visual Computer\\
VI & Visual Informatics\\
  CHI & ACM Conf. on Human Factors in Computing Systems\\
  EuroVis & Eurographics Conference on Visualization\\
  PacificVis & IEEE Pacific Visualization Symposium \\
  SIGGRAPH & ACM SIG on Comp. Graphics and Inter. Techniques\\
  VIS  & IEEE Visualization Conference\\
  \hline
BE & Building and Environment\\
CEUS & Computers, Environment and Urban Systems\\
EPB & Env. and Planning B: Urban Analytics and City Science\\
IJAC & Int. Journal of Architectural Computing\\
IJGIS & Int. Journal of Geographical Information Science\\
\highlight{P\&RS} & \highlight{Journal of Photogrammetry and Remote Sensing}\\
\highlight{ISPRS Ann.} &  \highlight{ISPRS Ann. of the Phot. Rem. Sens. and Spat. Inf. Sci.}\\
JUD & Journal of Urban Design\\
LUP & Landscape and Urban Planning\\
SCS & Sustainable Cities and Society\\
SimAUD & Symp. on Sim. for Architecture and Urban Design\\
UC & Urban Climate\\
  \hline
\end{tabular}
\vspace{-0.5cm}
\label{table:venues}
\end{table}

\section{Survey scope \& methodology}
\label{sec:methodology}

For our survey, we have focused our efforts on the selection of papers that (1) made visualization contributions leveraging 3D urban data or facilitating 3D urban visual analytics or (2) made domain-specific contributions generating or analyzing 3D urban data.
%
%Hence, our survey covers not only visual analytics tools, visualization techniques, and design and evaluation studies but also domain-specific papers that generate or analyze 3D urban data through modeling and simulations.
%
\highlight{Hence, our survey covers not only visual analytics tools, visualization techniques, and design and evaluation studies but also domain-specific papers that either analyze or generate 3D urban data. We have focused on the generation of thematic data that is either inherently 3D (e.g., CFD simulation of wind) or derived from the 3D physical layer (e.g., sunlight access). We have deliberately excluded the generation or acquisition of data for the physical layer as these works typically do not contain analytical components or thematic layers which are essential for 3D urban analytics.}

\highlight{The main goal of our survey was to review active areas of visualization research related to visual analysis of 3D urban data. For this reason, we reviewed papers from both visualization and domain-specific venues published between 2008 and 2023. While we acknowledge that there could be previous works published prior to 2008, focusing on the last 15 years helps manage the scope of the review while including studies that are more technologically and methodologically relevant.}
On the visualization side, we included the main visualization \& computer graphics journals (e.g., IEEE Transactions on Visualization and Computer Graphics) and conferences (e.g., IEEE VIS, EuroVis).
%
% \nivan{For the selection of domain-specific venues we collaborated with three domain experts (also co-authors of this survey) that are researchers in three different fields, namely urban planning, urban climate and geo-spatial data science.
% %
% They indicated the top journals and conferences in their areas that publish issues related to 3D urban data analytics.
% %
% }
%
In domain-specific venues, top journals (e.g., Computers, Environment and Urban Systems, Urban Climate) as well as a popular symposium (SimAUD), were included.
\begin{figure}[t!]
\begin{center}
    \includegraphics[width=\linewidth]{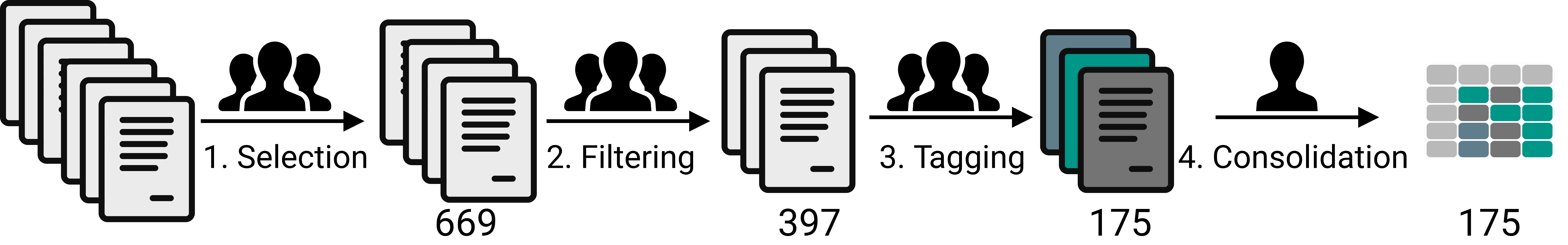}
\end{center}
% \vspace{-0.5cm}
\caption{In the selection, filtering, and tagging steps, each paper was reviewed by at least two of the authors. In the consolidation step, each paper was reviewed by one of the authors. We also highlight the number of papers considered at each step.}
\label{fig:methodology}
\end{figure}
These venues were selected based on discussions with experts in the field (also co-authors of this survey) and represent the top-ranked journals and conferences in GIS, architecture, \highlight{remote sensing} and urban planning that contain papers related to 3D urban visual/data analytics.
Table~\ref{table:venues} lists the initial set of visualization and domain-specific venues used in our search.

\begin{figure*}[ht!]
\begin{center}
\includegraphics[width=1\linewidth]{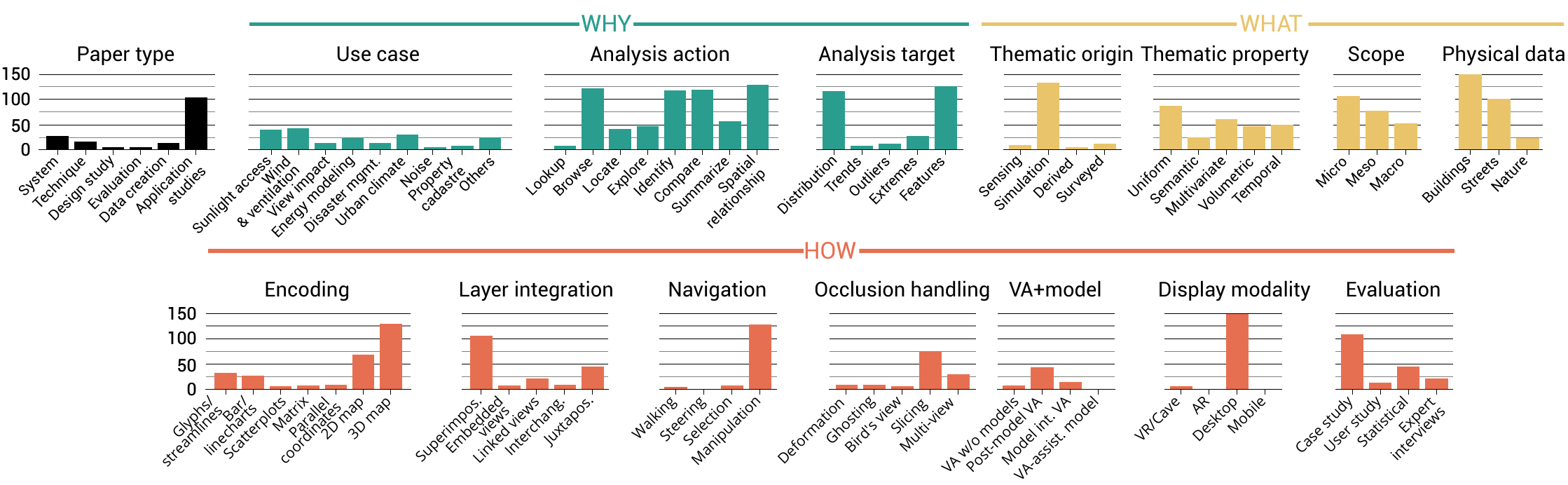}
\end{center}
% \vspace{-0.5cm}
\caption{Distribution of tags across \paper{Paper type}, \why{Why}, \what{What} and \how{How} dimensions.}
\label{fig:distributions}
% \vspace{-0.5cm}
\end{figure*}

To select relevant papers for the survey, we followed a four-step approach (Figure~\ref{fig:methodology}). In Step 1, we did an exhaustive search through the journals and proceedings highlighted in Table~\ref{table:venues}. In this step, each author was responsible for reviewing papers between 2008 and 2023 from 3 venues and selecting the ones with components in either (1) 3D urban data or (2) visual analytics for 3D urban data. 
\highlight{To achieve this we inspected the title, abstract and figures present in the paper (that could suggest the use and/or analysis of 3D urban data).}
This step resulted in an initial set of 669 papers.
In Step 2, we filtered the papers according to their relevance -- i.e., papers with either 3D urban data or visual analytics for 3D urban data. Each paper was randomly assigned to two authors of the survey; each survey author then independently assessed the relevance of the papers assigned to them. For each paper, if two survey authors agreed on the relevance of the paper, the paper was included; if not, the paper was removed. In case of a disagreement, a third survey author made the final decision.
\highlight{In this step, the authors followed a similar procedure as in the previous step, going through the title, abstract, and figures of each paper.}
A total of 397 papers remained after this step.

% an exhaustive search through journals and proceedings, followed by tagging, and then consolidation of the results (see Figure~\ref{fig:methodology}).
%
% In the first step, each author was responsible for reviewing papers between 2008 and 2023 from 3 venues and selecting the ones with components in either (1) 3D urban data or (2) visual analytics for 3D urban data. This step resulted in an initial set of 669 papers.
% \nivan{if relevant, we could detail a bit which pieces of information we collected from the papers during the process? While the process was subjective in a sense, we could try to detail the criteria used, by mentioning for example (in case that exists) what data from the papers were collected}
%
% Next, each author independently assessed the relevance of the identified papers, and papers classified as relevant by the \emph{majority} were selected for inclusion in the tagging stage -- i.e., papers with either 3D urban data or visual analytics for 3D urban data.
%

% \nivan{can we add one more sentence here stating what do we mean by relevant}\fabio{papers that satisfy the previously mentioned (1) or (2)}

In Step 3, we held meetings to discuss a set of informative tags based on the previously identified papers. Our intention was to have tags that reasonably describe each paper along three primary dimensions (\emph{why}, \emph{what}, and \emph{how}), covering use cases, analysis tasks, data, visualization, and interaction.
Each paper was then randomly assigned to two authors of the survey, who were responsible for individually classifying the paper using the tags.
Since the process of tagging demanded a more careful review of the papers, survey authors could also mark a paper as being outside of the scope of the survey. Following the same procedure outlined in Step 2, if there was disagreement regarding the removal of a paper, a third survey author made the final decision.
At the end of this step, out of 397 papers, 175 remained -- the other papers were excluded since, after a more careful inspection, they did not satisfy the aforementioned requirements.

% \nivan{and these were discarded?}\fabio{yes} \thomas{in this context R2 was interested in which keywords were kept and removed}

In Step 4, one author was responsible for consolidating the classifications of each paper into a final classification. At the end of this step, we had the final 175 tagged papers, including 54 visualization papers and 121 domain-specific ones.
%
% \nivan{this next phrase seems vague, Can we say why we included and what do we do with this?} \thomas{I was also wondering where and how they where included}
% We have also included the following commercial and open-source tools: deck.gl~\cite{wang_deckgl_2017}, UrbanSim~\cite{noth_urbansim_2003}, GRASS GIS~\cite{neteler_grass_2012}, ArcGIS~\cite{johnston_arcgis_2001}, kepler.gl~\cite{kepler_2023}, QGIS~\cite{qgis_2023} and Mapbox~\cite{mapbox_2023}.
%
These are works where both physical and thematic layers play a key role. In other words, works that leverage 3D geometries representing physical layers and data associated with such geometries defining thematic layers. As such, we have excluded works using urban data where there is no direct association between these layers (such as images) or where there is no clear presence of thematic attributes (such as point clouds).
For example, works that primarily focus on \emph{extracting} polygon meshes from images or point clouds are outside the scope of this survey.

% \nivan{maybe a version of this should also be present in the introduction}

\noindent \textbf{Limitations.} 
Given the multidisciplinary field, urban analytics found its way into a wide range of journals and conferences across different fields. 
% a myriad of journals and conferences have in their aims and scope topics related to urban sciences. 
Therefore, it is unfeasible for this survey to cover all the possible venues that might publish papers related to urban studies. 
Hence, early on, we decided to focus on two main types of venues: visualization ones and cross-cutting multidisciplinary venues that have, in their aims, topics related to urban problems and cities. Venues that do not satisfy these requirements include very specific (e.g.,~\emph{Journal of Wind Engineering and Industrial Aerodynamics}) to multidisciplinary ones (e.g.,~\emph{Nature}).
In doing so, we were able to restrict the number of surveyed venues while maintaining a reasonable sample of use cases, analysis workflows, and visualization techniques. We believe that, while not an exhaustive search through \emph{every} journal and conference that might publish on urban studies, we present a well-defined picture of the topic to bridge the gap between disciplines.

\section{Survey overview and organization}
\label{sec:overview}

In this survey, we summarize previous visualization and domain-specific contributions using an interrogative method that tries to answer: \why{Why} is 3D urban data being analyzed, \what{What} data is being analyzed, and finally \how{How} it is being analyzed.
%
% We also frame these questions around primary, data, and visualization \& interaction dimensions.
%
% \nivan{a similar phrase appeared not long before this one. Maybe we should remove it:} \thomas{probably unclear what are `primary dimensions' in this context}
Given the interdisciplinary nature of urban analytics, our goal was to cover both visualization aspects of the surveyed works (to allow us to identify research gaps more easily) as well as urban domain aspects (to facilitate the creation of a common language between visualization researchers and urban experts, beyond siloed collaborations).
% \thomas{lot of relevant bracketed info makes it hard to grasp, split sentence?}

Figure~\ref{fig:overview} presents an overview of the survey, with the three main questions, their categories, and fine-grained tags.
A paper could be assigned to one or more tags within a given category, except for paper type, where each paper is only assigned to one type.
We dedicate a separate section to each one of these questions/dimensions.
Section~\ref{sec:paper_types} presents the considered paper types in this survey; Section~\ref{sec:why} then presents the \why{Why} dimensions, namely use cases (\ref{sec:usecase}), analysis actions (\ref{sec:action}) and targets (\ref{sec:target}); Section~\ref{sec:what} presents the \what{What} dimensions, that is physical data entities (\ref{sec:what-physical}), thematic data origin (\ref{sec:what-thematic-origin}) and properties (\ref{sec:what-thematic-properties}), and spatial scope (\ref{sec:what-scope}). Then in Section~\ref{sec:tasks}, we provide an overview of the most common tags from the previous sections in order to characterize a set of popular tasks across the use cases. Section~\ref{sec:how} discusses the \how{How} dimensions, which include visualization encodings (\ref{sec:how-encoding}), integration of thematic and physical visualizations (\ref{sec:how-integration}), occlusion handling (\ref{sec:how-occlusion}), navigation (\ref{sec:how-navigation}), integration (\ref{sec:how-integration}), display modalities (\ref{sec:how-technology}), and evaluation methods (\ref{sec:how-evaluation}). 
Each section starts with a brief overview of that specific dimension and its corresponding categories. Each subsection then drills down into various aspects of that domain, data and visualization \& interaction, with plots showing the distribution of tags in that category. For easier referencing, we \paper{highlight} the first occurrence of the tag in the text. 
% \thomas{this is not 100\% true, should we just always highlight tags to make them more visible?}
% 
% Section~\ref{sec:takeaways} presents the takeaways from the survey, Section~\ref{sec:challenges} presents research directions and opportunities, and Section~\ref{sec:conclusion} concludes the survey.
%
Section~\ref{sec:new_challenges} presents research directions and opportunities, and Section~\ref{sec:conclusion} concludes the survey.

Figure~\ref{fig:distributions} shows a summary of the number of papers with each tag. In the image, tag popularity can be compared across dimensions. 
Figures~\ref{fig:how-why}, \ref{fig:how-what}, \ref{fig:why-what} provide an overview of the distribution of tags, but now as the intersection across \why{Why}, \what{What}, and \how{How} dimensions.
To facilitate the exploration of the distribution considering the intersection of tags, the color of each cell in the matrices is normalized by the total number of papers within each sub-matrix.
For example, the color of each cell in the \emph{Use case} \& \emph{Paper type} sub-matrix (Figure~\ref{fig:how-why} (top left)) is normalized by the total number of papers within that sub-matrix.

% . The color of each cell in the matrix is normalized by the total number of papers in that combination of categories \nivan{maybe we could include a quick sentence explaining the rationale behind this normalization. It might help to explore the distribution for the combination of tags, something like that}. . \thomas{since we are providing an example here, I think we should also provide the actual total number. Also Figure~\ref{fig:how-why} is rather far away, maybe top of next page or switching with Figure~\ref{fig:distributions} as it is more relevant}

\begin{figure*}[t!]
\begin{center}
\includegraphics[width=\linewidth]{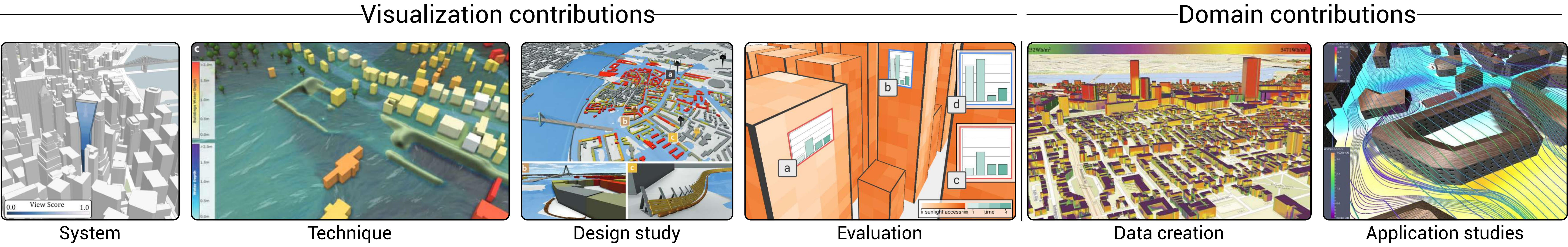}
% \vspace{-0.5cm}
\caption{Example of papers from each one of the types considered in the survey. Left: papers with visualization contributions (system~\cite{doraiswamy_topology-based_2015}, technique~\cite{cornel_interactive_2019}, design study~\cite{waser_many_2014}, evaluation~\cite{mota_comparison_2022}). Right: papers with domain contributions (data creation~\cite{liang_visualization-oriented_2014} and application studies~\cite{koch_compact_2018}).}
\label{fig:papertypes}
\end{center}
% \vspace{-0.5cm}
\end{figure*}

\section{Paper type}
\label{sec:paper_types}

One initial effective way to understand the span of contributions in 3D urban data analytics is to classify the research works and their paper types. We followed the categories proposed by Munzner~\cite{munzner_process_2008} and classified visualization papers using the following categories: \paper{system}, \paper{technique}, \paper{design study}, and \paper{evaluation}. \highlight{We note that Munzner's model paper type was not used, since none of the surveyed visualization papers aligned with its classification.}
While the previous classification is well suited for visualization and visual analytics works, it does not cover the range of possible contributions in domain-specific papers. To account for these papers, we have added two new paper types: \paper{data creation}  and \paper{application studies}.
%
% \begin{wrapfigure}{r}{0.3\textwidth}
% \begin{center}
% \includesvg[inkscapelatex=false, width=\linewidth]{figs/paper-type.vl.svg}
% % \includegraphics[width=\linewidth]{figs/Untitled-4.pdf}
% \end{center}
% % \caption{\label{fig:type_distribution}Distribution of  paper types. \marcos{Some bar charts have too much white space in the horizontal direction and very small labels. Maybe reducing the white space may allow making some of them taller while keeping their width.}}
% \end{wrapfigure}
%
The \paper{data creation} type refers to domain-specific works that present methodologies to derive novel pieces of \highlight{thematic} data that are not available or are very difficult to acquire through other methods. 
% that can not be feasibly acquired by other methods. 
Similarly, the \paper{application studies} type refers to domain-specific papers presenting detailed analytical studies using 3D urban data. 
We highlight that for these two paper types, the data visualization component is not a contribution, but rather often a tool to inspect and present their results. For example, while design studies propose visual representations for a particular domain problem, application studies contribute domain-specific analysis of 3D urban data.
%
% \thomas{R3 could not make out the difference between design study and application study, check if this is sufficiently clear, maybe give an example of an application study with reference}
%
%
Figure~\ref{fig:papertypes} shows examples of this categorization, with visualization and domain contributions.
Each of the surveyed papers was assigned to one of the paper types, resulting in the distribution presented on the top left of Figure~\ref{fig:distributions}.
%
% \nivan{maybe remove: The common thread linking all of these papers is the necessity to analyze 3D urban data, as previously defined.}

\section{Primary dimensions (\why{Why})}
\label{sec:why}
% \thomas{what does `primary dimensions' mean in context of `Why:'. I am not sure how we came up with this heading. probably unclear for reader.}

\begin{figure*}[t!]
\begin{center}
\includegraphics[width=0.93\linewidth]{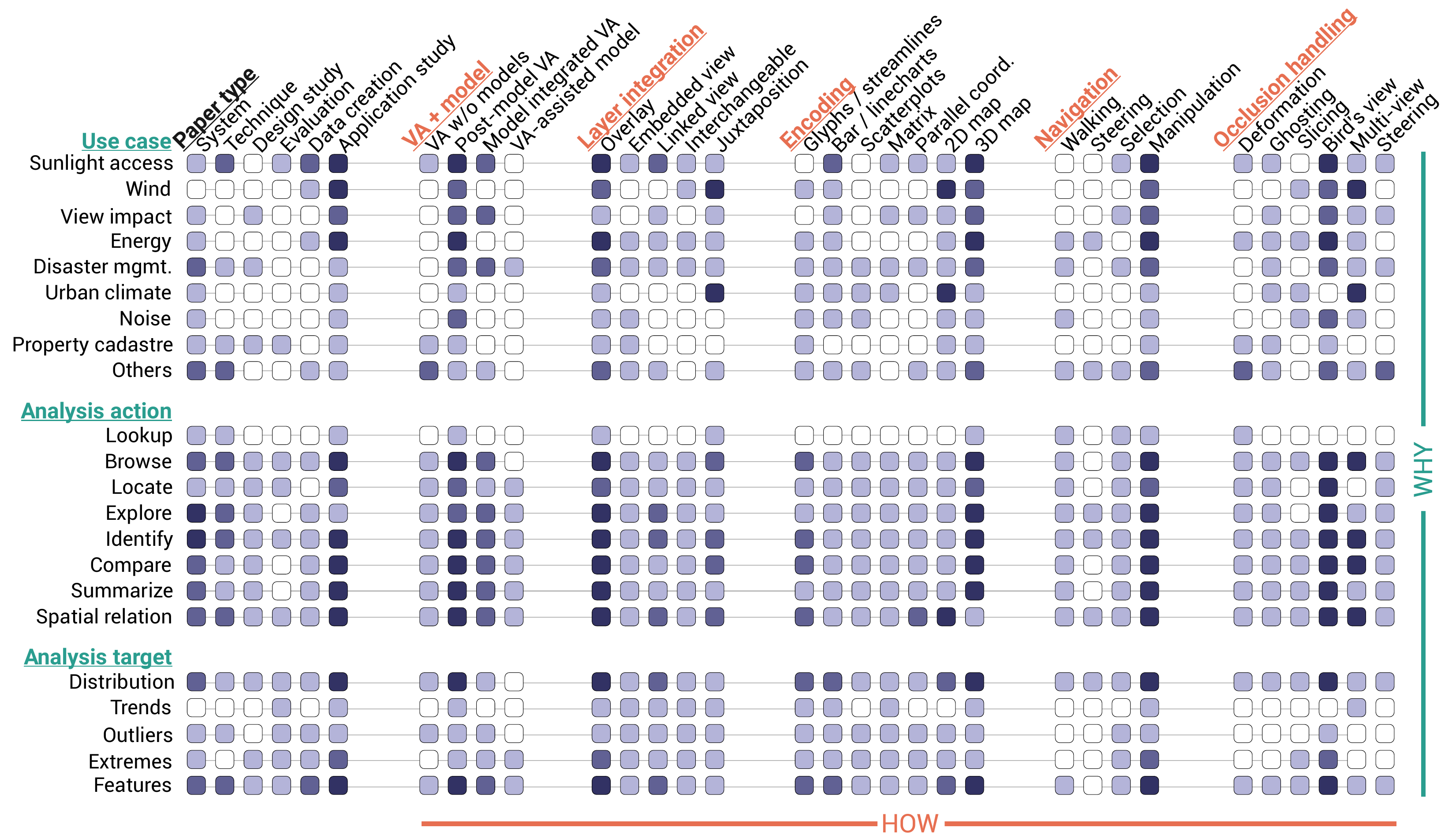}
% \vspace{-0.25cm}
\caption{Distribution of surveyed papers according to \why{Why} and \how{How} dimensions, with shades denoting tag occurrence.}
\label{fig:how-why}
\end{center}
% \vspace{-0.75cm}
\end{figure*}

% Given the 
% overwhelming challenges facing cities today, 
As metropolitan areas continue to expand and adopt more complex forms, they face new challenges that are large in scope and scale. Facilitated by the proliferation of new sources of data, urban analytics emerged to address those challenges by probing and analyzing cities.
%
% \nivan{this phrase on interdisciplinarity appears over and over again} \marcos{the previous text is also a repetition of ideas that appears in the abstract and introduction}
% Urban analytics is an interdisciplinary field encompassing various disciplines, such as architecture, urban planning, engineering, and computer science.
%
At its core is the need to integrate, model, analyze, and visualize multiple sources of data describing different urban phenomena while taking into account the requirements and expectations of multiple stakeholders, from subject matter experts to concerned citizens.
%
% For this reason,
% \thomas{in the following two sentences I think it is unclear what we mean and what we did}
In our primary dimensions, we discuss papers taking into account \why{Why} 3D urban data is being analyzed. 
%
% we establish classifications that go towards answering the question of \emph{Why} is 3D urban data being analyzed.
%
First, we present the result of the classification of the papers considering their primary domain \emph{use case} (Section~\ref{sec:usecase}). Then, we review the papers according to their \emph{analysis action} (Section~\ref{sec:action}) and \emph{analysis target} (Section~\ref{sec:target}).
Figure~\ref{fig:how-why} shows the distribution of tags according to the \why{Why} and \how{How} dimensions.

% Points to mention: 
% %
% (1) System / technique / design study / evaluation primarily cover visualization papers (categorized according to visualization domain).
% %
% (2) Data / analysis primarily cover domain papers.
% %
% (3) Each paper was only classified in one category.
% %
% (4) As an interdisciplinary topic, mention balance between vis papers and domain papers.

% \subsection{Urban data life cycle}

% \begin{wrapfigure}{r}{0.4\textwidth}
% \begin{center}
% \includesvg[inkscapelatex=false, width=\linewidth]{figs/cycle.vl.svg}
% \end{center}
% \end{wrapfigure}

% Points to mention:
% %
% (1) Explain urban data pipeline.
% %
% (2) Talk about data interoperability, straddling different systems.
% %
% (3) Lack of curation and transformation works, even thought it has been mentioned in several papers.

\subsection{Use cases}
\label{sec:usecase}

% \begin{wrapfigure}{r}{0.35\textwidth}
% \begin{center}
% \includesvg[inkscapelatex=false, width=\linewidth]{figs/use-cases.vl.svg}
% \end{center}
% \end{wrapfigure}

The survey covers a wide breadth of urban use cases, each with different data, analytics, and visualization requirements.
We classified each paper into the following use cases: sunlight access, wind, view impact, building energy modeling, disaster management, urban climate, noise, property cadastre, and others. 
Such classification was decided after our initial selection process (detailed in Section~\ref{sec:methodology}) and it comprehensively covers all surveyed works. Use cases with fewer than five papers are classified as ``others.''
To facilitate the discussion in this section, the use cases are grouped into three broader themes: (1) contributions that model or analyze the interplay between an inherently 3D natural phenomenon and the built environment (Section~\ref{sec:natural}); (2) contributions that model or analyze 3D phenomena that are primarily driven by human factors (Section~\ref{sec:human}); and (3) contributions that are in and of itself defined by the built environment (Section~\ref{sec:built}).

\subsubsection{3D natural phenomenon and the built environment}
\label{sec:natural}

In the first theme, we have papers that cover sunlight access, wind \& ventilation, damage \& disaster management, and urban climate.
\why{Sunlight access} analysis studies the impact of the built environment (primarily buildings) on the equitable ``right to light'' or ``right to sunshine'', with a wide range of impacts, from vegetation growth
to solar energy potential and building energy consumption. 
The majority of surveyed papers related to sunlight access were domain-specific ones (25 out of 39), either proposing new efficient approaches to compute data (Figure~\ref{fig:papertypes}~(Data creation)), studies analyzing the energy potential in different cities~\cite{zhang_study_2013, bremer_new_2016, delmas_solar_2016, kaynak_software_2018, de_luca_computational_2017, ouria_evaluation_2018, zhu_solar_2019, florio_designing_2021} or for different scenarios, such as pedestrian comfort~\cite{de_luca_sun_2019}, photovoltaic panels~\cite{fijalkowska_assessment_2022, zhu_economically_2022,zheng_feasibility_2023}, urban farming~\cite{palliwal_3d_2021}, retrofit~\cite{saretta_integrated_2020} or data standards~\cite{biljecki_variants_2016}.
While not their primary target, a number of domain-specific papers use sunlight access and shading as one of their attributes of interest~\cite{albeaik_virtual_2017, yao_3dcitydb_2018, wilson_how_2019, wolosiuk_application_2020}.
Regarding visualization contributions tackling sunlight access, the majority were interactive visualization systems specifically designed for this type of problem~\cite{miranda_shadow_2019} or that use sunlight as an attribute of interest~\cite{wu_virtual_2010, engel_immersive_2012, willenborg_applications_2017, zhang_urbanvr_2021}.
Doraiswamy~et~al., for instance, present Catalogue, a system that combines different view measures to create building design options (Figure~\ref{fig:papertypes} (System))~\cite{doraiswamy_topology-based_2015}.
Shadow Profiler is a tool aimed at urban planners and architects driven by the need to assess shadows at a city scale (Figure~\ref{fig:shadow}~(left))~\cite{miranda_shadow_2019}.
Technique contributions directly linked with sunlight include Mu\~{n}oz et al. with real-time computation of solar exposure~\cite{munoz-pandiella_real-time_2017}; 
Bremer et al. propose a sampling and texture mapping approach for solar potential simulations~\cite{bremer_new_2016}.
Indirectly, Pasewaldt et al. use sunlight access as a case for their multi-perspective detail+overview contribution~\cite{pasewaldt_multi-perspective_2013}; Vanegas et al. propose an inverse design procedure for quick creation of 3D models~\cite{vanegas_inverse_2012}; and Lorenz and D\"{o}llner use sunlight access as a case for their discussion on mapping 3D building geometry to 2D rasters~\cite{lorenz_3d_2010}.
Lastly, Herbert and Chen~\cite{herbert_comparison_2015} and Mota et al. present evaluation studies, respectively comparing 2D and 3D representations for urban planning and 3D-oriented visualizations for urban analytics (Figure~\ref{fig:papertypes}~(Evaluation))~\cite{mota_comparison_2022}.

All of the surveyed papers about \why{wind \& ventilation} were domain-specific ones (43).
Wilkinson et al. present a data generation technique to approximate complex computational fluid dynamics (CFD) using an artificial neural network\cite{wilkinson_approximating_2014}. The other surveyed papers were primarily analysis contributions that relayed on CFD for building-level~\cite{avini_wind_2019} or neighborhood-level~\cite{elshaer_variations_2017, zhang_cfd_2021} wind load analysis (Figure~\ref{fig:physicalthematic} (Superimposition)), wind power potential~\cite{zabarjad_shiraz_wind_2020}, inner building ventilation~\cite{shi_evaluating_2019, hadavi_quantifying_2020}, wind impact on urban climate~\cite{gros_cool_2014,de_luca_sun_2019,hadavi_impacts_2021}, pollution dispersion~\cite{gallagher_parked_2019,zhu_influence_2022, he_effects_2023,li_cfd_2023}, or the interplay between sunlight access and wind for thermal comfort~\cite{johansson_wind_2020}.

\why{Damage \& disaster management} was a use case where we had more visualization contributions than domain-specific ones (9 out of 13). Here, we included papers that tackled heavy rain and floods, seismic response, and search \& rescue.
Analysis papers include Xiong et al.~\cite{xiong_building_2015} and Willemborg et al.~\cite{willenborg_applications_2017} with applications of 3D urban models on building seismic response and simulation of detonations.
When considering the visualization papers, two present design studies targeting visual analytics for flood management~\cite{waser_many_2014, konev_run_2014} (Figure~\ref{fig:papertypes} (Design study)). Cornel et al. present a technique for surface reconstruction with real-world applications in flood and heavy rain scenarios (Figure~\ref{fig:papertypes} (Technique))~\cite{cornel_interactive_2019}.
Other surveyed papers propose visual analytics systems~\cite{ribicic_visual_2013,chen_application_2014,cornel_visualization_2015,bock_visualization-based_2017,vuckovic_combining_2021, li_3d_2019,boorboor_submerse_2023}. Vuckovic et al., for instance, explore the combination of 2D and 3D visualizations for flood and stormwater management~\cite{vuckovic_combining_2021}.
At the intersection between several use cases (i.e., sunlight access, wind \& ventilation, and damage \& disaster management use cases), Elsayed investigates the interplay between multiple 3D variables, such as solar radiation, ventilation, and air temperature, for pandemic mitigation plans~\cite{salaheldin_microclimatic_2021}.

A large number of papers in the category of 3D natural phenomena primarily tackle problems related to \why{urban climate} (30 in total). Most papers (28) were application studies, investigating surface temperature~\cite{hu_analysis_2019}, and the impact of urban morphology~\cite{karimimoshaver_effect_2021} and redevelopment~\cite{koch_compact_2018} on climate~(Figure~\ref{fig:papertypes}~(Application studies)).
Other contributions include analysis based on WRF simulations~\cite{kadaverugu_improving_2021,wong_integrated_2021,huang_sensitivity_2023}.
Visualization contributions include Gautier et al.'s system to enable domain scientists to analyze the interplay between air temperature and the built environment~\cite{gautier_co-visualization_2020}, and Deng et al.'s context-aware technique to visualize street-level data~\cite{deng_interactive_2016}.

\subsubsection{3D phenomena driven by human factors}
\label{sec:human}

In this theme, we have contributions that model or analyze 3D phenomena that are primarily driven by human factors, such as energy modeling or energy potential assessment, noise \& sound propagation, and property cadastre.
\why{Urban building energy modeling} (UBEM) has been recognized as an increasingly useful data-driven approach to tackle the many challenges facing cities today and identify pathways for building retrofit and deployment of city-level renewable energy initiatives~\cite{abbasabadi_urban_2019, johari_urban_2020, ang_ubemio_2022}. 
Out of the 23 papers on the topic, 17 were classified as application study papers, either analyzing specific urban cases~\cite{ouria_evaluation_2018, fijalkowska_assessment_2022, zhu_economically_2022}, the interplay between sunlight and energy usage~\cite{saretta_integrated_2020}, or urban climate and energy usage~\cite{gros_cool_2014, nouvel_influence_2017,willenborg_applications_2017, hadavi_impacts_2021}. Other domain-specific contributions include general frameworks (with simple visualization components) for the assessment of energy usage~\cite{torabi_moghadam_gis-statistical_2018, abbasabadi_integrated_2019, ang_ubemio_2022}.
Data creation contributions include Albeaik et al. with the creation of 3D urban models for energy and sunlight assessment~\cite{albeaik_virtual_2017}, and Krietemeyer and Kontar~\cite{krietemeyer_method_2019} and Wolosiuk and Mahdavi~\cite{wolosiuk_application_2020} with data integration methods for building performance assessment.
Bartosh and Gu presented an immersive visualization system that enabled users to explore energy consumption data~\cite{bartosh_immersive_2019}.

\why{Noise \& sound propagation} is another use case of interest when considering 3D urban data. In recent decades, noise has become a growing urban problem, impacting public health~\cite{hammer_environmental_2014}, social well-being~\cite{guite_impact_2006} and quality of life~\cite{dratva_impact_2010}, as noise increases stress, sleep disruption, annoyance and distraction~\cite{bronzaft_neighborhood_2007,haralabidis_acute_2008}.
Noise codes typically impose regulations that aim to mitigate the noise at the source (e.g., by erecting sound barriers or modifying building designs)~\cite{bronzaft_invisible_2010,hammer_environmental_2014}, adding new requirements for urban planners and architects to optimize designs, which in turn creates the need to analyze noise and sound propagation.
Out of the eight surveyed works on this topic, six were domain-specific, with analyses exploring the relationship between the built environment and noise~\cite{stoter_3d_2008,zhang_case_2017,tang_dynamic_2022}. The only visualization contributions were an immersive system for the visualization of traffic noise~\cite{bartosh_immersive_2019}, and a visualization design for continuous noise phenomena, with a case study aimed at urban noise~\cite{beran_third_2022}.

Visualizations in 3D also play an important role in the \why{property cadastre} case, especially in the last few years with the increase in the demand for 3D real property cadastre~\cite{jazayeri_geometric_2014,drobez_transition_2017}. Unlike rural areas, where a 2D description of the land is sufficient, urban areas present several challenges, where property units (and land uses) are positioned on top of each other. In our survey, we reviewed eight papers on this topic. 
Application studies include Koziatek and Dragi\'{c}evi\'{c}'s iCity3D, a forecasting method for vertical urban development~\cite{koziatek_icity_2017}.
Surveyed visualization papers also include two evaluations on how users perceive transparency~\cite{wang_how_2017} and rendering attributes~\cite{seipel_visualization_2020} in visualizations of 3D cadastre; a design study discussing the requirements for 3D cadastre systems~\cite{shojaei_visualization_2013}; a visualization prototype for 3D cadastre~\cite{shojaei_design_2018}; a distortion technique for 3D building properties~\cite{ying_distortion_2019}; and an immersive system that included property management~\cite{lv_managing_2016}.

\subsubsection{Built environment only}
\label{sec:built}

In this theme, we have contributions where the built environment itself is the major focus of analysis. This theme covers \why{view impact analysis}, a common operation in architecture and urban planning. Scores computed on the surface of the building summarizing the visibility of certain geographical features (e.g., landmarks, parks, waterfronts).
Our survey reviewed 14 papers in this theme. Analysis papers tackled view access equity~\cite{yasumoto_use_2011}, visibility to green spaces~\cite{yu_view-based_2016,virtanen_near_2021} and landmarks~\cite{salimi_visual_2023}, views of high rises~\cite{li_room_2022} and office rents~\cite{turan_development_2021}.
The visualization papers include: Ortner et al., a design study for visibility-aware urban planning~\cite{ortner_vis--ware_2017}; Zhang et al., an immersive analytics system that takes into account visibility~\cite{zhang_urbanvr_2021}; Doraiswamy et al., a topology-based framework for building design~\cite{doraiswamy_topology-based_2015}; and Ferreira et al., with a visual analytics system aimed at architects and urban planners that allows for view impact analyses~\cite{ferreira_urbane_2015}.

Other use cases in this theme include the analysis of urban change over time~\cite{butkiewicz_multi-focused_2008,koziatek_local_2019}, walkability considering 3D footpath networks~\cite{sun_connecting_2021,zhao_walkability_2021}, simulation of radio propagation in city environments~\cite{bounceur_cupcarbon_2018}, enclosure assessment~\cite{kaya_modelling_2017}, urban design plans~\cite{leidi_exploring_2013,reinhard_urban_2015,miao_computational_2018,levine_spatial_2022,sun_generative_2023}, and the study of urban vitality~\cite{zeng_vitalvizor_2018} (Figure~\ref{fig:physicalthematic} (Linked view)).

% First, contributions that tackle the interplay between an inherently 3D natural phenomenon and the built environment, which is primarily represented by high-resolution building data. In this category, we have papers that cover the analysis of sunlight access (which primarily depends on the interplay between sunlight and the built environment), wind \& ventilation (movement of gases and the built environment), and climate (temperature and the built environment at high spatial resolution).

% Second, contributions that model or analyze 3D phenomena that are primarily driven by human factors. In this category, we have contributions that visualize the output or inspect building energy models, damage \& disaster management, noise \& sound propagation and property cadastre.

% Finally, in the third category we have contributions that are in and of itself defined by the built environment, without any exogenous factors playing a significant role. This category covers view impact analysis, a common operation in architecture and urban planning where, at certain positions in a building, scores are computed summarizing the visibility to certain geographical features (e.g., landmarks, parks, water fronts).

% Points to mention:
% %
% (1) Breadth of topics.
% %
% (2) Each topic has different visualization and data needs.
% %
% (3) Mention the topics covered in others.

\subsection {Analysis actions}
\label{sec:action}

% \begin{wrapfigure}{r}{0.3\textwidth}
% \begin{center}
% \includesvg[inkscapelatex=false, width=\linewidth]{figs/analysis-action.vl.svg}
% \end{center}
% \end{wrapfigure}

In order to evaluate the surveyed papers on the reasons \emph{why} they analyze 3D urban data, we follow Brehmer and Munzner and their typology of abstract visualization tasks~\cite{brehmer_multi-level_2013,munzner_visualization_2015} that include mid-level (search) and low-level (query) actions. Given that the surveyed works are driven by the discovery and analysis of new information (as opposed to presentation), we decided to focus only on search and query actions.
Given the importance of spatial awareness and spatial relation, we incorporate a spatial awareness query action, following Elmqvist and Tsigas' taxonomy of 3D occlusion management for visualization~\cite{elmqvist_taxonomy_2008}. The term \emph{spatial relation} will be used in the remainder of this survey.

\subsubsection{Search}

\begin{figure}[t!]
\begin{center}
\vspace{0.5cm}
\includegraphics[width=\linewidth]{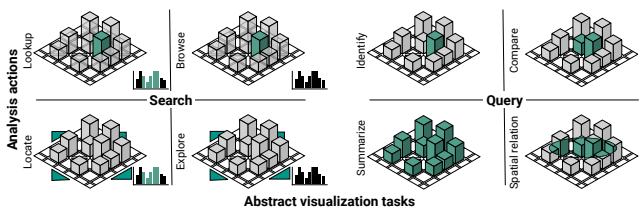}
\caption{Search (left) and query (right) actions.}
\label{fig:searchquery}
\end{center}
% \vspace{-0.75cm}
\end{figure}

Brehmer and Munzner classify search actions according to whether the location and identity of targets of analysis are known to the user.
When considering 3D urban data, the physical layer contains the location of the analysis, and the thematic layer contains the identification, i.e., the aspect of the data that the user is interested in.
Figure~\ref{fig:searchquery} (left) highlights the four possible cases, with a known location represented as a green building \location, and a known identity represented as a bar chart with selected values~\target. An unknown location is represented by four directional arrows~\nolocation, and an unknown identity is represented by a bar chart without any values selected~\notarget.

A \why{lookup} search is one where the user knows both the location~\location and the identity~\target. For example, a user performing view impact analysis might want to look up the areas of a building with the best views of parks only considering the north-facing fa\c{c}ade.
Wu et al. present a framework that allows users to perform lookup-like sunlight analysis for a known building at a specific time of the year (Figure~\ref{fig:scales} (micro))~\cite{wu_virtual_2010}.

% Deng et al.\cite{deng_interactive_2016} presented a method for disocclusion for urban environments that balances both disocclusion and distortion; their proposal highlights a potential use case with pollution data where the user knows both the location (street segments within a region)

In a \why{locate} search, the user knows the identity~\target, but does not know the location of the target, requiring them to look around different places of the physical layer~\nolocation. For example, a user analyzing UBEM might want to locate the apartment with the highest energy consumption in a certain neighborhood.
The domain papers by Yu et al.~\cite{yu_view-based_2016}, and Li et al.~\cite{li_room_2022} perform a locate search for view impact analysis (Figure~\ref{fig:scales} (middle, bottom)). In the papers, the analysis is interested in locating places in a fa\c{c}ade that have high scores for certain points of interest (e.g., greenery, water body).
Locate search has also been highlighted by Huang et al. and Wickramathilaka et al. when visualizing areas exceeding a particular wind speed~\cite{huang_sensitivity_2023} and noise~\cite{wickramathilaka_three_2023} threshold, respectively.

A \why{browse} search is one where the user knows the location~\location but does not know the identity~\notarget of the target. In property cadastre, for example, the user might be interested in searching the number of ownership changes of units in a particular building.
As shown in Figure~\ref{fig:how-why}, the browse search is the most popular type among the domain papers, which is expected given the non-interactive (i.e., batch-oriented) type of analyses performed in these studies, with limited exploration capabilities.
This type of search is also commonly supported by distortion techniques, such as Wu et al.~\cite{wu_multiperspective_2016} and Chen et al.~\cite{chen_urbanrama_2022} (Figure~\ref{fig:oclusion} (Deformation)), as the user knows the location and distortion is used to enable the identification of the target.
The UrbanVR visualization system presented by Zhang et al.~\cite{zhang_urbanvr_2021} is an example of how the analysis of 3D urban data can be supported by a browse search (Figure~\ref{fig:oclusion} (Assisted)): the user is interested in the analysis of different candidate buildings for a particular location, but needs to evaluate against certain attributes (e.g., shading and visibility).

In an \why{explore} search, the user does not know the location~\nolocation nor the identity~\notarget of the target, usually starting with an overview of the data at a macro scale. In sunlight access analysis, for example, a user might be interested in exploring temporal patterns across all buildings of a neighborhood.
This type of search is more commonly supported by visual analytics systems, as shown in Figure~\ref{fig:how-why}. Tools like Run Watchers~\cite{konev_run_2014}, Urbane~\cite{ferreira_urbane_2015}, Cornel et al.~\cite{cornel_visualization_2015}, Vis-A-Ware~\cite{ortner_vis--ware_2017}, VitalVizor~\cite{zeng_vitalvizor_2018}, Shadow Profiler~\cite{miranda_shadow_2019}, and UrbanVR~\cite{zhang_urbanvr_2021} all support exploration as part of their analyses. Even though certain domain-specific papers support exploration~\cite{yao_3dcitydb_2018,tang_dynamic_2022}, they mostly rely on off-the-shelf 3D visualization libraries, with very little effort going towards the customization of these interfaces to satisfy specific tasks.
An outlier is the work by Reinhard that combines 2D and 3D visualizations for the exploration of urban planning projects~\cite{reinhard_urban_2015}.

\subsubsection{Query}

% \begin{wrapfigure}{l}{0.29\textwidth}
% \begin{center}
% \includegraphics[width=\linewidth]{figs/query.pdf}
% \caption{Query actions.}
% \label{fig:query}
% \end{center}
% \end{wrapfigure}

Following Brehmer and Munzner's query actions, after a target (or set of targets) is found by the user through search actions (either on the physical or thematic layers), the user will then \emph{identify}, \emph{compare}, or \emph{summarize} these targets. To account for the need to have spatial awareness, we also classify papers according to Elmqvist and Tsigas' idea of \emph{spatial relation}. Figure~\ref{fig:searchquery} (right) presents the four possible cases.
While an \emph{identify} query refers to a single target~\location, \emph{compare} refers to multiple subsets of targets~\location\location, and \emph{summarize} will require an overview of a large subset of the data. 
% \thomas{how does `spatial relation' compare regarding the number of targets?}

In the context of 3D urban data, an \why{identify} query is one where the primary target of interest is a single region of the fa\c{c}ade, an entire fa\c{c}ade, or an entire building. For example, in sunlight access analyses, a user might be interested in identifying floors or windows with the largest amount of accumulated shade over a period of time.
Most of the surveyed papers (both domain and visualization ones) support this query. For example, for the identification of radio propagation values~\cite{bounceur_cupcarbon_2018}, ideal locations for wind turbines in cities~\cite{zabarjad_shiraz_wind_2020}, discrepancies between simulated and predicted wind values~\cite{wilkinson_approximating_2014}, and energy consumption in historical neighborhoods~\cite{torabi_moghadam_gis-statistical_2018}.

In a \why{compare} query, the user might be interested in comparing different targets -- for example, comparing sunlight access of north and south-facing fa\c{c}ades or comparing between different \emph{what if} scenarios.
Delmas et al. compare different building configurations to maximize solar potential in a neighborhood~\cite{delmas_solar_2016}.
Figure~\ref{fig:scales} presents two analysis cases where comparisons are at the center of the analysis workflow~\cite{li_room_2022, yu_view-based_2016}. In Figure~\ref{fig:scales} (middle), the analysis compares two different view impact attributes across buildings in a neighborhood (middle-center) and across a fa\c{c}ade (middle-right). Similarly, in Figure~\ref{fig:scales} (bottom-right), the view access to greenery is compared across three different sides of the building, while (bottom-center) performs a \emph{what-if}  analysis (or scenario planning) comparing view access before and after the placement of a tall neighboring building.

In a \why{summarize} query, the user is interested in analyzing a large set of targets -- for example, an overview of the sunlight access situation across buildings, parks, roads, and sidewalks in a dense neighborhood.
Stoter et al. present an overview of the noise impact on a 3D urban environment~\cite{stoter_3d_2008}, highlighting limitations of 2D noise maps, and De Luca summarizes wind flows for the assessment of pedestrian comfort at a neighborhood scale~\cite{de_luca_sun_2019}.

Lastly, in a \why{spatial relation} query, the user is interested in the relation of spatial properties of a target and its context (whether it is in the physical or thematic layers).
Assessing this particular query is especially difficult, given the lack of analysis details in some papers. Therefore, for this classification, we looked for the use of keywords such as ``context'', ``neighborhood'', and ``vicinity'' in the description of the requirements, methodology, or analysis.
A large number of surveyed papers (129) could then be classified as performing this query.
The topology-based system proposed by Doraiswamy et al., for instance, highlights the need for \emph{3D context models} and \emph{explore the view extent over the city}~\cite{doraiswamy_topology-based_2015}, while the study by Moghadam et al. assesses the concentration of energy consumption over neighborhoods~\cite{torabi_moghadam_gis-statistical_2018}.
Spatial relation is also important in wind analyses~\cite{elshaer_variations_2017, avini_wind_2019}, given the impact of neighboring buildings on the aerodynamics of a tall building.

\subsection{Analysis targets}
\label{sec:target}

Lastly, we classified papers according to their primary abstract targets. We included five targets covered by Munzner~\cite{munzner_visualization_2015}: trends, features, extremes, distributions, and outliers. The target of analysis was considered with respect to the thematic layer.
%
% , i.e., some aspect of the data of interest to the user \thomas{this sentence is rather vague and probably falls into the criticism of R3. maybe we can relate this to thematic data explained earlier}. . Certain targets from Munzner's taxonomy were excluded, either because they are common to all papers (e.g., spatial shape), or are present in fewer than five papers (e.g., network topology).
%
% \begin{wrapfigure}{r}{0.3\textwidth}
% \begin{center}
% \includesvg[inkscapelatex=false, width=\linewidth]{figs/analysis-target.vl.svg}
% \end{center}
% \end{wrapfigure}
%
A \why{trend} target is a pattern in the data, e.g., the high energy consumption of apartment units during summer.
For example, Nouvel et al. compare the impact of data quality of 3D buildings on urban heating demand modeling, analyzing trends over different city districts~\cite{nouvel_influence_2017}.
A \why{feature} target is any structure in the data that might be of interest in the analysis task, e.g., summer energy consumption within certain blocks of a city.
Chen et al. propose an exploded view technique for immersive urban analytics in which the feature target is the road network occluded by buildings~\cite{chen_immersive_2017}.

An \why{extreme} target is the minimum or maximum value when considering a specific range, e.g., maximum summer energy consumption for apartments in a building.
Cornel et al. propose an uncertainty-aware visualization of vulnerability that includes extreme values across flooding scenarios~\cite{cornel_visualization_2015}.
Hu and Wendel evaluate minimum and maximum values in their anisotropy analyses over two neighborhoods in New York City, leveraging 3D building data~\cite{hu_analysis_2019}.
Delmas et al. optimize building placement to maximize solar potential~\cite{delmas_solar_2016}, and De Luca and Voll analyze daily minimum sunlight hours~\cite{de_luca_computational_2017}.
View impact analyses such as the paper by Yu et al. are also interested in extreme targets to identify floors with the least occluded views of landmarks~\cite{yu_view-based_2016}.
A \why{distribution} target is one where the user is interested in the spread of values considering a range, e.g., distribution of energy consumption for apartments in a building.
The papers classified as systems usually support both exploration and summarization, therefore enabling the visualization of spatial~\cite{li_room_2022} and spatiotemporal~\cite{miranda_shadow_2019, mota_comparison_2022} distributions.
Finally, an \why{outlier} target is the one that deviates from that trend, e.g., a specific apartment with low energy consumption during hot summer days.
Even though no paper explicitly mentions outlier as their analysis target, the target is supported by many of the visual encodings used by the surveyed papers, as we detail in Section~\ref{sec:how-encoding}.

\begin{figure*}[t]
\begin{center}
\includegraphics[width=0.95\linewidth]{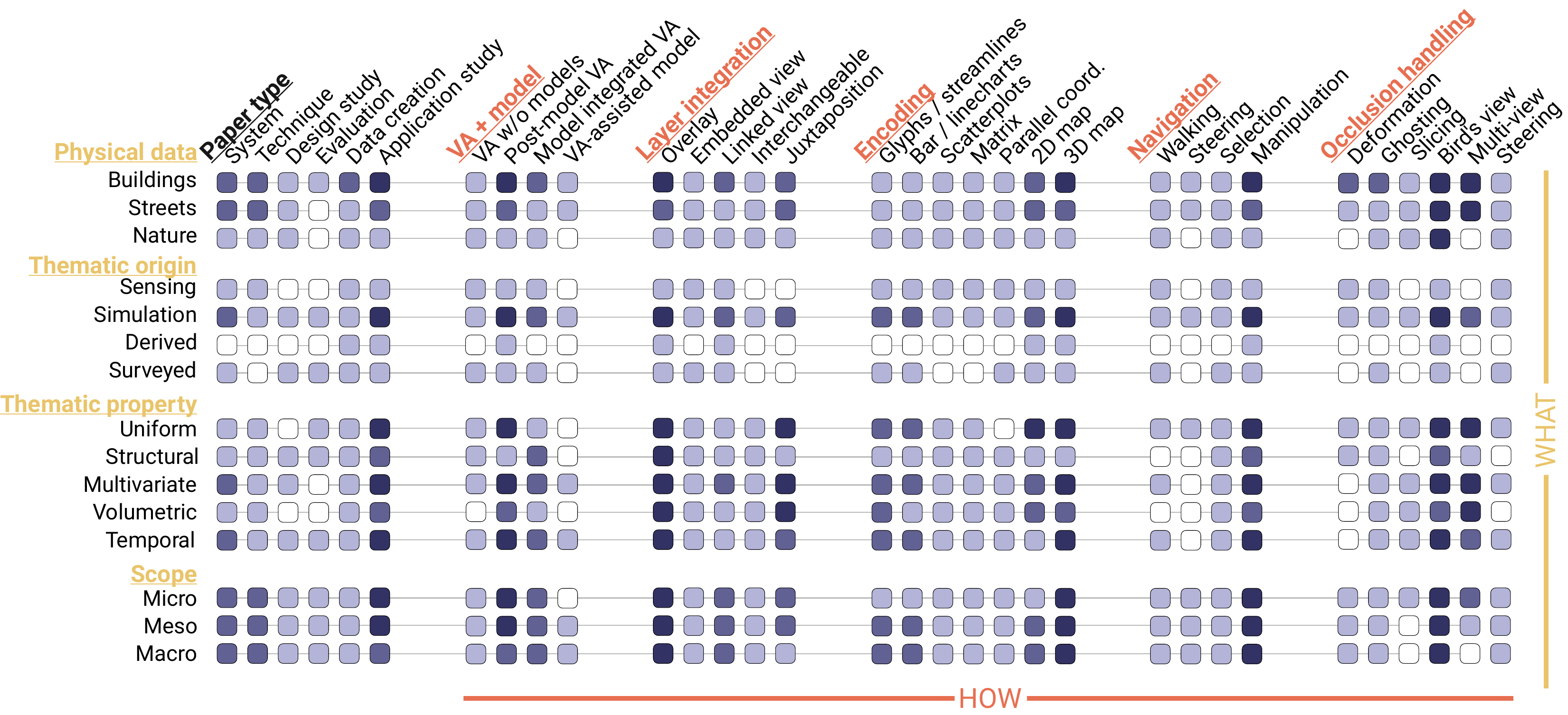}
% \vspace{-0.25cm}
\caption{Distribution of surveyed papers according to \what{What} and \how{How} dimensions, with shades denoting tag occurrence.}
\label{fig:how-what}
\end{center}
% \vspace{-0.75cm}
\end{figure*}

%%%%%%%%%%%%%%%%%%%%%%%%%%%%%%%%%%%%%%%%%%
\section{Data dimensions (\what{What})}
\label{sec:what}
%%%%%%%%%%%%%%%%%%%%%%%%%%%%%%%%%%%%%%%%%%
%

% \nivan{It is a small thing, but to me it makes more sense to talk about the physical data first. It is the most basic one and the one we talk the last, maybe we should get over it sooner. Also, we probably should add something about the type of data we are talking abour here meshes lidar and maybe a few challenges in dealing with them.}

The 3D datasets used for the different use cases covered in this survey impose several challenges to all stages of the urban data life cycle. From modeling 3D geometries and deploying physical sensors to developing computational strategies capable of sampling inherently 3D urban-related phenomena and aspects, the collection/generation and management of 3D data is either expensive, computationally intensive, or both. 
Also, 3D data is provided in a myriad of different formats and resolutions depending on their sources, type of information, and analysis requirements. For this reason, curating, managing, analyzing, and visualizing these data require specifically designed data solutions.

In this section, we will then highlight \what{What} datasets and properties were used in the surveyed works.
Given the aforementioned challenges and the tight connection between visual analytics and data, in this survey, we identified four data dimensions to categorize the reviewed papers:
to highlight the connection between physical and thematic layers, we have included a category with the physical entities considered in the data -- i.e., buildings, streets, nature (Section~\ref{sec:what-physical});
the source of the domain-specific 3D thematic data (Section~\ref{sec:what-thematic-origin});
the properties of such data where we cover whether the data is uniform, multivariate, volumetric, temporal, and has an attached structural meaning (Section~\ref{sec:what-thematic-properties});
and the urban spatial coverage of the data -- i.e., micro, meso, macro (Section~\ref{sec:what-scope}).
Figure~\ref{fig:how-what} shows the distribution of tags according to the \what{What} and \how{How} dimensions.

%%%%%%%%%%%%%%%%%%%%%%%%%%%%%%%%%%%%%%%%%%
\subsection{Physical data entities}
\label{sec:what-physical}
%%%%%%%%%%%%%%%%%%%%%%%%%%%%%%%%%%%%%%%%%%
% \begin{wrapfigure}{r}{0.3\textwidth}
% \begin{center}
% \includesvg[inkscapelatex=false, width=\linewidth]{figs/physical-data-entities.vl.svg}
% \end{center}
% \end{wrapfigure}

\begin{figure}[b!]
\begin{center}
\includegraphics[width=1\linewidth]{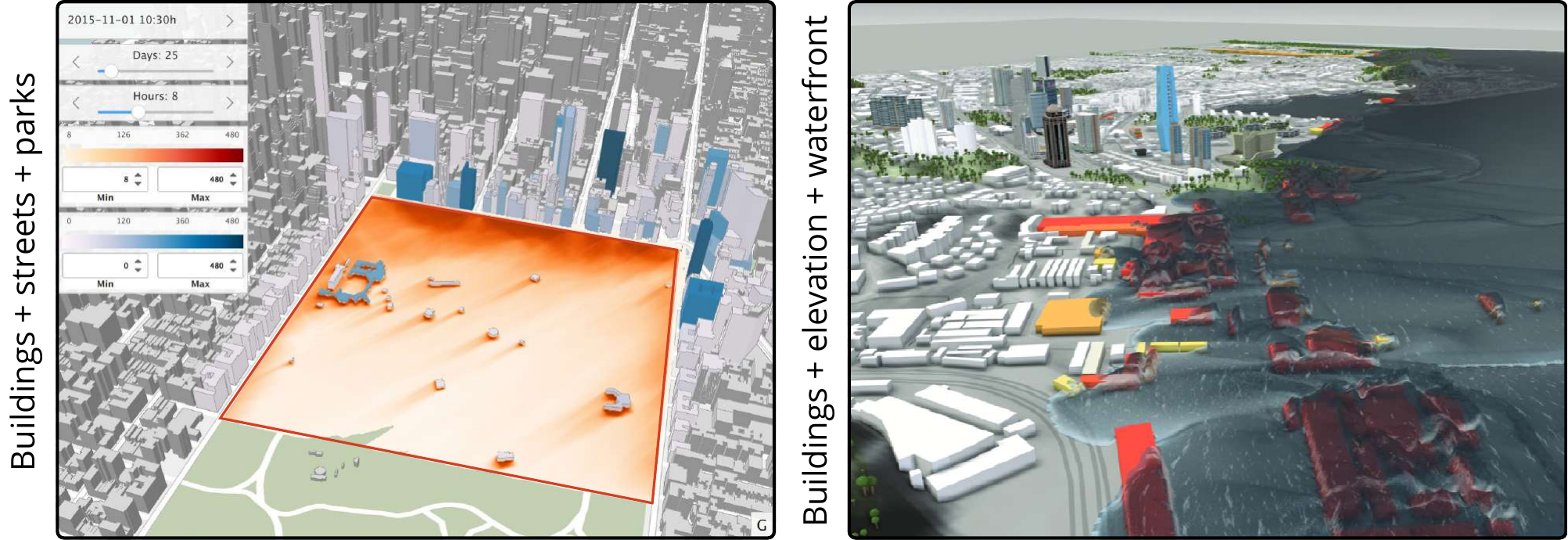}
\caption{Examples of integration between physical and thematic layers for sunlight access analysis (left, \cite{miranda_shadow_2019}) and disaster mgmt. (right, \cite{cornel_interactive_2019}).}
\label{fig:shadow}
\end{center}
% \vspace{-0.75cm}
\end{figure}

%
In conjunction with thematic data, 3D urban data papers also utilize data describing the built environment -- called here physical data.
Physical data is fundamental since it gives geometric support and spatial context to the thematic data.
Depending on the resolution of both the physical and thematic data, the latter must be aggregated over the former.
In our survey, we have considered three types of physical data entities: \what{buildings}, \what{streets}, and \what{nature}, with nature encompassing parks, trees, mountains, hills, and waterfronts.
Figure~\ref{fig:shadow} shows an example of visual analytics tools making use of multiple physical layers and aggregating shadow contributions at a building level.
The topography of the urban environment is particularly important for papers tackling disaster \& damage management. Cornel et al.~\cite{cornel_visualization_2015, cornel_interactive_2019}, Vuckovic et al.~\cite{vuckovic_combining_2021}, and Waser et al.~\cite{waser_many_2014} all make use of elevation data for both modeling and visualization purposes.

% \thomas{for me this paragraph feels a bit out of place. We talk about data sources, data formats and data types. This does not look like an exhaustive listing, so we need to justify why we present this subset}

In the surveyed works, OpenStreetMap~\cite{haklay_open_2008} is the primary source of data for buildings, streets, and certain nature-related features, such as parks and waterfronts.
OpenStreetMap makes data available as a series of crowdsourced tags attached to data structures representing nodes, ways, and relations. Buildings, for example, are defined as 2D outlines with associated heights (or number of floors) and roof types, requiring adequate data processing to extract polygon meshes for visualization.
Alternatively, authoritative sources usually make building data available using the CityGML standard with geometry information specified at a certain level of detail, or directly as polygon meshes. Government agencies are the primary source for tree data, usually made available as tabular data with the location of trees.
For topographic features, in our survey, we have found that works usually make use of terrain elevation rasters made available by OpenTopography and/or NASA's Shuttle Radar Topography Mission.
Point cloud and image data are also potential sources for data describing the built environment, though they require advanced techniques to generate polygon meshes and attributes to be leveraged by visual analytics tools~\cite{biljecki_street_2021, hu_towards_2021, hu_sensaturban_2022}.

% 3D visual analytics papers utilize traditional geographical information system data (called here physical data). Physical data is important since it gives geometric support and context to the thematic data. In fact, depending on the resolution of both the physical and thematic data, the latter must be aggregated over the former. In~\cite{,} although thematic data is provided based on a uniform subdivision of the building facade, aggregated information is also visualized per building to give an overview to the user. Also, some applications require the analysis of multiple physical entities, like buildings and street networks~\cite{,}, for example.

%%%%%%%%%%%%%%%%%%%%%%%%%%%%%%%%%%%%%%%%%%
\subsection{Thematic data origin}
\label{sec:what-thematic-origin}
%%%%%%%%%%%%%%%%%%%%%%%%%%%%%%%%%%%%%%%%%%
% \begin{wrapfigure}{r}{0.3\textwidth}
% \begin{center}
% \includesvg[inkscapelatex=false, width=\linewidth]{figs/thematic-data-origin.vl.svg}
% \end{center}
% \end{wrapfigure}

The thematic data considered in the surveyed works were classified according to their origin, i.e., how they were acquired or generated. 
Given that many of the considered use cases depend on modeling and simulation approaches, such as view impact, sunlight access, wind, and seismic analysis, it is no surprise that most works relied on \what{simulation}-based datasets.
In this classification, we have also included data created from mathematical models, such as CFD~\cite{allegrini_coupled_2017,llaguno_shaping_2018,nazarian_numerical_2018,allegrini_simulations_2018,karimimoshaver_effect_2021,li_cfd_2023}, WRF~\cite{wong_integrated_2021,kadaverugu_improving_2021,huang_sensitivity_2023}, and ENVI-met~\cite{haeri_evaluation_2023}.

The second most popular data origin was data acquired using \what{sensing} methods. In this category, we include remote sensing~\cite{lorenz_3d_2010,catita_extending_2014,bonczak_large-scale_2019}, data gathered by robots~\cite{bock_visualization-based_2017} or by sensors deployed in the urban environment~\cite{ouria_evaluation_2018,morrison_urban_2021}. Palliwal et al., for example, leverage sensors placed along the fa\c{c}ade of a building to estimate urban farming potential~\cite{palliwal_3d_2021}.

% \thomas{I am not sure if surveyed might be confusing to some readers, because drones and robots are also used for surveying tasks.}

The third data origin category in our work covers data that come from authoritative sources and/or \what{surveyed} data, such as land use description~\cite{ferreira_urbane_2015, koziatek_icity_2017, zeng_vitalvizor_2018}, demographic data~\cite{chang_legible_2007}, public infrastructure information~\cite{bartosh_immersive_2019}, or property cadastre~\cite{seipel_visualization_2020}. We have found that survey data is usually used as a criterion to filter the data in certain visualizations. For example, both Urbane and VitalVizor have land use attributes as dimensions in their brushable parallel coordinates.

Finally, we have also considered approaches that \what{derived} new data from raw data using computational models, including machine learning ones. Examples include approaches for urban energy modeling~\cite{abbasabadi_integrated_2019}, wind flow~\cite{mokhtar_conditional_2020,kabosova_fast_2022}, and view impact~\cite{li_room_2022}.
It is interesting to point out that we observed a growing number of learning-based 3D data generation approaches in the last five years, which is expected given the omnipresence of neural network-based solutions nowadays.

%%%%%%%%%%%%%%%%%%%%%%%%%%%%%%%%%%%%%%%%%%
\subsection{Thematic data properties}
\label{sec:what-thematic-properties}

%%%%%%%%%%%%%%%%%%%%%%%%%%%%%%%%%%%%%%%%%%
% \begin{wrapfigure}{r}{0.35\textwidth}
% \begin{center}
% \includesvg[inkscapelatex=false, width=\linewidth]{figs/thematic-data-properties.vl.svg}
% \end{center}
% \end{wrapfigure}

In our survey, we found that thematic data has distinct spatiotemporal characteristics.
Regarding their spatial information, 3D thematic datasets can be categorized into two broad categories: (1) data defined at the surface of physical data entities (e.g., buildings) and (2) data defined inside 3D volumetric regions, not necessarily inside buildings.
We can further subdivide category (1) into two subcategories: data uniformly discretized on the surface of physical data entities (e.g., solar radiation exposure on different parts of the building), not necessarily aligning with real-world structures like windows and floors (Figure~\ref{fig:papertypes} (data creation)), and data discretized into smaller units based on physical aspects with semantic meaning, e.g., structural elements such as apartment units, windows, floors, massings (Figure~\ref{fig:papertypes} (System)). In our classification, we use \what{uniform} and \what{structural} to refer to these two categories, respectively.
The choice between uniform and structural discretization is based on the use case of the paper. For example, energy modeling~\cite{ang_ubemio_2022} and 3D property cadastre~\cite{shojaei_visualization_2013} associate data with individual buildings, floors, or property units. Therefore a structural discretization is more appropriate.
Out of the eight property cadastre surveyed works, four make use of structural discretization.
On the other hand, seismic simulations~\cite{xiong_building_2015} primarily rely on a uniform discretization along the building's surface.
Some other use cases, such as sunlight access and view impact analysis, make use of both uniform (e.g., view impact computed on the surface of the building)~\cite{doraiswamy_topology-based_2015, miranda_shadow_2019} and structural discretizations (e.g., view impact computed for each building unit)~\cite{ferreira_urbane_2015, turan_development_2021}.
In category (2), data defined inside 3D \what{volumetric} regions, we have data produced by simulations and/or related to natural phenomena (Figure~\ref{fig:papertypes} (Application studies)), such as flood and heavy rain~\cite{cornel_interactive_2019}, wind~\cite{hadavi_quantifying_2020, zabarjad_shiraz_wind_2020, zhang_cfd_2021} and sound or radio propagation~\cite{bounceur_cupcarbon_2018, tang_dynamic_2022}.
Such data is particularly popular in papers leveraging CFD wind data, with 74\% making use of volumetric data.

We also classified papers on whether they used \what{temporal} data. Since urban regions are dynamic and affected by human behaviors, natural phenomena, and other elements that are strongly associated with periods of the day, days of the week, and seasons of the year, it was no surprise to discover that many of the selected papers consider thematic datasets containing a temporal component. For example, Mota et al. present an evaluation of different designs for visual analytics tasks considering sunlight data for the different seasons of the year~\cite{mota_comparison_2022}.

Finally, we classified the thematic data based on whether they make use of \what{multivariate} data. View impact analysis is an example of a use case that takes into account several attributes to compute visibility scores (e.g., view scores to greenery and landmarks). Turan et al., for instance, take into account four different attributes in their view impact analysis~\cite{turan_development_2021}. 
Visualization systems~\cite{ferreira_urbane_2015,ortner_vis--ware_2017,miranda_shadow_2019,zhang_urbanvr_2021} usually support multidimensional datasets through standard or new visualization encodings. For example, Waser et al. propose a novel interface for the visualization of multidimensional ensembles targeting the problem of flood management~\cite{waser_many_2014}.

%%%%%%%%%%%%%%%%%%%%%%%%%%%%%%%%%%%%%%%%%%
\subsection{Spatial data scopes}
\label{sec:what-scope}
% \marcos{should we add an image with examples like images \ref{fig:physicalthematic} and \ref{fig:oclusion}?}
%%%%%%%%%%%%%%%%%%%%%%%%%%%%%%%%%%%%%%%%%%
%
% \begin{wrapfigure}{r}{0.3\textwidth}
% \begin{center}
% \includesvg[inkscapelatex=false, width=\linewidth]{figs/analysis-scope.vl.svg}
% \end{center}
% \end{wrapfigure}

\begin{figure}[t!]
\begin{center}
\includegraphics[width=\linewidth]{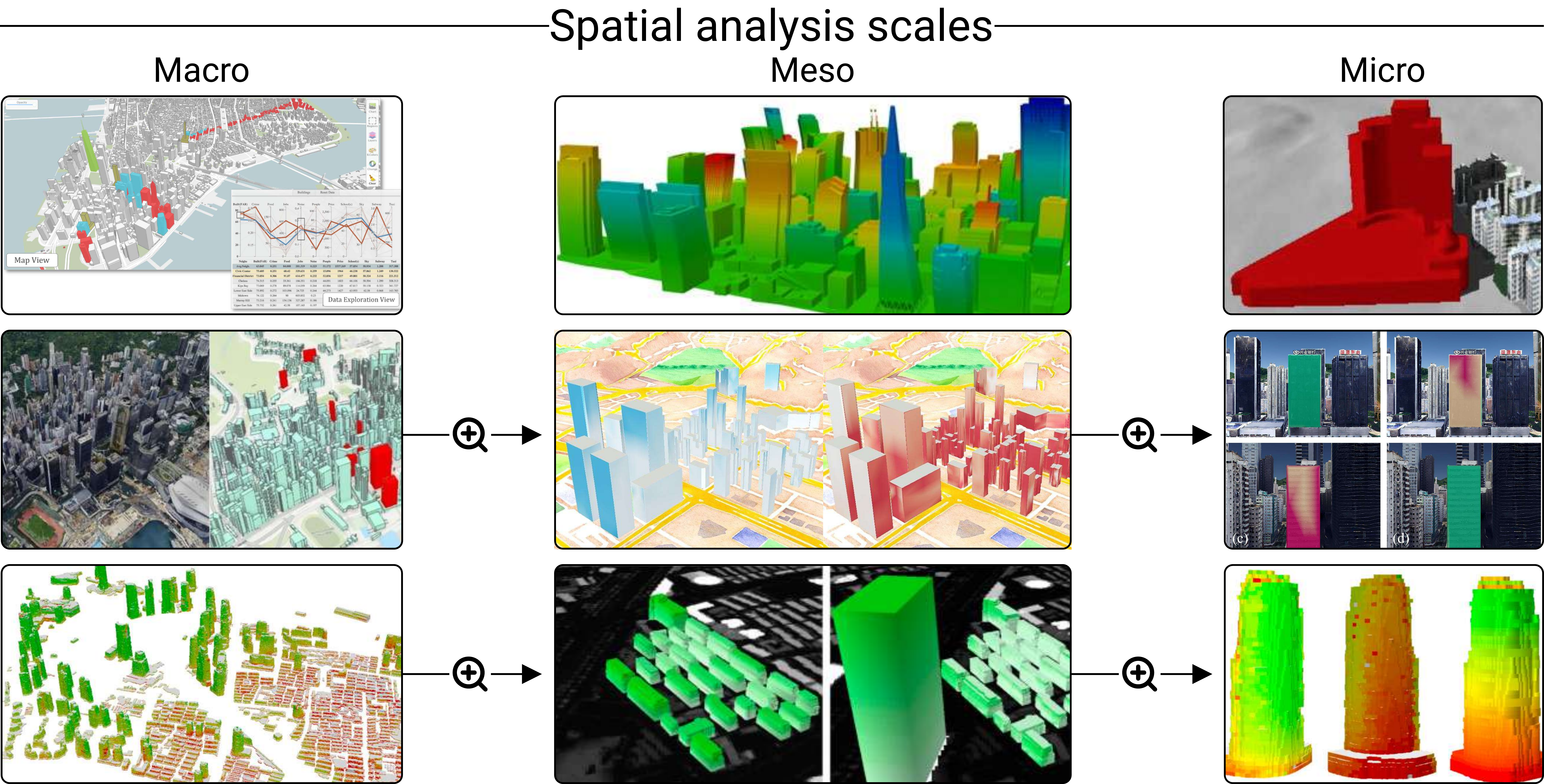}
\caption{Examples of different analysis scopes. Top left: view impact analysis at a city scale~\cite{ferreira_urbane_2015}. Top center: seismic analysis at a neighborhood scale~\cite{xiong_building_2015}. Top right: shadow impact analysis at a building scale~\cite{wu_virtual_2010}. Middle and bottom: two domain papers with view impact analysis across the three scales~\cite{li_room_2022, yu_view-based_2016}.}
\label{fig:scales}
\end{center}
% \vspace{-1cm}
\end{figure}

The surveyed papers focused on three different spatial resolutions for their analytical tasks.
We have identified approaches that consider specific sites (e.g., fa\c{c}ade~\cite{palliwal_3d_2021} or buildings~\cite{pasewaldt_multi-perspective_2013, de_luca_computational_2017}), multiple sites (e.g., one or several blocks~\cite{zhang_study_2013, munoz-pandiella_real-time_2017, kaynak_software_2018, karimimoshaver_effect_2021, li_room_2022, zhu_economically_2022}), or large regions (several neighborhoods or the entire city~\cite{ferreira_urbane_2015, miranda_shadow_2019, chen_urbanrama_2022}). We call these approaches \what{micro}, \what{meso}, or \what{macro} scale, respectively. Figure~\ref{fig:scales} (top) presents three examples of analyses performed at different scales: view impact using Urbane at the macro scale~\cite{ferreira_urbane_2015}, seismic analysis by Xi et al. at the meso scale~\cite{xiong_building_2015}, and sunlight access analysis by Wu et al. at the micro scale~\cite{wu_virtual_2010}.

It is interesting to notice that analysis across multiple scales is performed in a number of domain papers. Figure~\ref{fig:scales} (middle, bottom) shows a view impact analysis performed across the three different scales~\cite{li_room_2022, yu_view-based_2016}. In these analyses, the domain expert is not only interested in an individual building but also in the interaction between multiple neighboring buildings and the spatial distribution over different neighborhoods.

% papers handling 3D urban data enable the visualization and exploration of thematic data at different spatial resolutions. We identified approaches that consider specific sites~\cite{,}, single block~\cite{,}, multiple blocks~\cite{,}, and the entire city~\cite{,}. It is interesting to notice that only a handful of the surveyed papers propose visual analytical approaches that consider all these four resolutions~\cite{,}.

% It is worth noticing that the spatial analysis scope is also highly related to the paper's use cases. In fact, papers tackling disaster management usually focus on specific locations of the city and are more likely to consider lower resolutions~\cite{,} \marcos{is this true?}. On the other hand, papers containing climate use cases consider higher resolutions, usually citywide visualizations~\cite{,}. \marcos{Is this also true? I think that a What x Why matrix may be necessary to improve the discussion over the entire WHAT section.}

% Points to mention: (1) Certain papers straddle different resolutions (make note of the handful that cover all five resolutions). (2) Drill down and mention the distribution of resolutions by use case (e.g., disaster management usually cover what resolutions?).

\section{Task characterization}
\label{sec:tasks}

\begin{figure}[t!]
\begin{center}
\includegraphics[width=\linewidth]{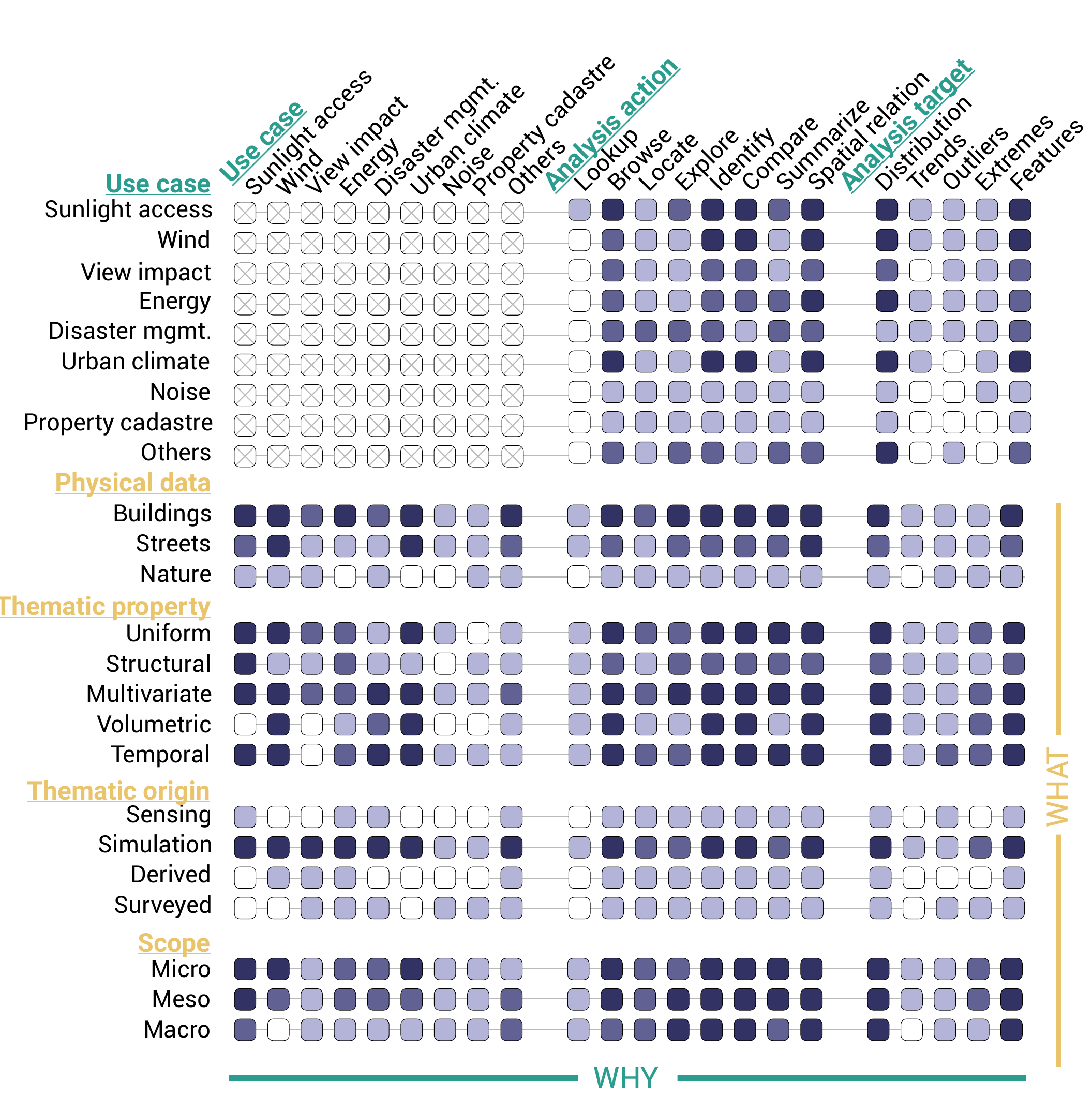}
% \vspace{-0.5cm}
\caption{Distribution of surveyed papers according to \what{What} and \why{Why} dimensions, with shades denoting tag occurrence.}
\label{fig:why-what}
\end{center}
\vspace{-0.5cm}
\end{figure}

Designing visualizations for 3D urban data involves the careful consideration of data and intended analyses. In this section, we use the distribution of \what{What} and \why{Why} tags to discuss relevant tasks across the different use cases considered in this survey (Figure~\ref{fig:why-what}).
\highlight{We begin with six guiding questions (Q1-Q6). Questions are on what are the most common search actions, query actions, analysis targets, thematic data properties, spatial scopes, and physical data entities. We then use the distribution of tags along use cases, i.e. the answers to Q1-Q6, to derive common visualization tasks. 
This methodology follows the approach of Mota et al.~\cite{mota_comparison_2022}, but using the distribution of tags within a use case to infer its commonness.}

%we answer the question about a tag to be common with yes when it occurs in more than 50\% of the reviewed papers.

\subsection{Guiding questions}

%We follow the approach presented by Mota et al., where task characterization is framed along six guiding questions (Q1-6)~\cite{mota_comparison_2022}.
%
\highlight{To answer questions Q1-Q6 about the occurrence of tags, we consider that a specific tag is common with respect to a use case if it is associated with more than 50\% of the surveyed papers of that case.}
At the end of each question, we summarize the most common use cases for each tag.
%
% For each question, we highlight \emph{primary} tags across use cases. We consider that a tag is relevant if two or more papers make use of it in their analysis workflow.

\noindent \textbf{Q1. What are the most common search actions?}

The majority of the surveyed studies analyze data at a specific location but without knowing the identity of the target -- i.e., a browse (\location+\notarget) search action. In our survey, we have found that such action is common in all considered use cases.
However, we found that a locate (\nolocation+\target) search action is only common in disaster management (77\% of the papers) and noise (50\%) use cases.
Similarly, an explore (\nolocation+\notarget) action is also common in disaster management (62\%) and noise (50\%) use cases.
In our review, we have identified fewer than ten works that make use of lookup search (\location+\target).

% all but the property cadastre case. In this case, an expert may be interested in evaluating values that match a specific value at different spatial locations.

% Domain specialists may be interested in analyzing data at a specific spatial location, but without knowing the identity of the target -- i.e., a browse (\location+~\notarget) search action. In our survey, we have found that such action is relevant in all considered use cases.
% %
% A locate (\nolocation+~\target) search action is relevant in all but the property cadastre case. In this case, an expert may be interested in evaluating values that match a specific value at different spatial locations.
% %
% In an explore action (\nolocation+~\notarget), relevant in all but wind and climate cases, an expert is evaluating data without any specific target.
% %

\noindent \why{Browse}: Common in all use cases; \why{Locate}: Disaster mgmt. and noise; \why{Explore}: Disaster mgmt. and noise.

\noindent \textbf{Q2. What are the most common query actions?}

We have found that identity and spatial relation actions are common in all use cases. For example, over 85\% of the surveyed papers tackling disaster management consider both identify and spatial relation tasks.
Among wind studies, all of the papers make use of spatial relation actions, given the fundamental role that the built environment plays in determining wind patterns.
The summarize action is common in four use cases: noise (63\%), energy (52\%), disaster management (54\%), and view impact analysis (50\%).
The compare action is common in all but the property cadastre case (only 25\% of the papers).

\noindent \why{Identify}: Common in all use cases; \why{Spatial relation}: Common in all use cases; \why{Compare}: Common in all except property mgmt; \why{Summarize}: View impact, energy, disaster mgmt., noise.

\noindent \textbf{Q3. What are the most common analysis targets?}

The analysis of features is common among all of the surveyed use cases, with more than 50\% of the papers in each category considering it.
The analysis of distributions is popular in all of the use cases, with the exception of property cadastre.
The analysis of extreme values is only common in noise studies (63\% of the papers in that category), and we found that analysis of trends and outliers is not common in any use case.

\noindent \why{Features}: Common in all use cases; \why{Distribution}: Common in all except property cadastre; \why{Extremes}: Noise.

\noindent \textbf{Q4. What are the most common thematic data properties?}

The presence of data uniformly discretized on the surface of physical objects is common in four different use cases: urban climate (90\%), noise (88\%), wind (88\%), and sunlight access (59\%) analysis.
Conversely, the use of data discretized at a structural level (e.g., floors) is only common in the property cadastre case, where data is associated with individual floors or property units.
In our survey, we have also found that multivariate data is common among urban climate (77\%), disaster management (73\%), and view impact (50\%) cases.
Temporal data is common in disaster management (85\%) and urban climate (53\%), and volumetric data is common in wind (74\%) and urban climate (53\%) cases.

\noindent \what{Uniform}: Urban climate, noise, wind, sunlight access; \what{Structural}: Property cadastre; \what{Multivariate}: Urban climate, disaster mgmt, view impact; \what{Temporal}: Disaster mgmt, urban climate; \what{Volumetric}: Wind, urban climate.

\noindent \textbf{Q5. What are the primary spatial scopes?}

We have found that micro-scale analyses are common in all of the surveyed use cases, particularly sunlight access (87\%), wind (81\%), and disaster management (77\%).
In the surveyed papers, mesoscale analyses are popular in all use cases, with the exception of wind and property cadastre.
Macro-scale analyses, on the other hand, are common in disaster management~(54\%).

\noindent \what{Micro}: Common in all use cases; \what{Meso}: Common in all except property cadastre; \what{Macro}: Disaster mgmt.

\noindent \textbf{Q6. What are the primary physical data entities?}

As expected, given the scope of the survey, building data is a common data entity across all use cases.
Street data is commonly leveraged in use cases studying urban climate (100\%), wind (88\%), noise (75\%), and disaster management (69\%).
On the other hand, nature data is not common in any use cases, being moderately popular only in disaster management (38\%) and view impact (29\%) cases.

\noindent \what{Buildings}: Common in all use cases; \what{Streets}: Urban climate, wind, noise, disaster mgmt.

\subsection{Common analysis tasks}

% \thomas{T5 and T6 seem to confuse R2 as they focus on volumetric data and uncertainty which could also play a role in T1-T4. Not sure how to address this. Maybe these are more `analysis patterns' than clear cut tasks? This would also avoid confusion with the tasks (analysis actions) described in the Why Section}

Common visualization tasks can be derived from the combination of the tags and their frequency that we derived in Q1-6 in the previous section.
Next, we highlight six of these common tasks across multiple use cases (\textbf{T1-6}). While not exhaustive, these tasks are diverse in their analytical (\why{Why}) and data (\what{What}) dimensions.
%
%For each task, we also highlight condensed responses (\textbf{R1-6}).
%
We also note that these tasks are oftentimes combined, e.g., analysis of volumetric data (T5) can be accompanied by uncertainty analysis (T6).
In subsequent sections, we highlight how different visualization techniques facilitate these tasks.

\noindent \textbf{T1. Micro-scale browsing for site identification and comparison}

A common task in urban use cases is the building or fa\c{c}ade-level analysis for the identification of potential sites matching certain requirements for equipment installation (e.g., solar panels) or urban redevelopment.
For example, urban planners are interested in the analysis of the distribution of sunlight access data where the general location is known (a single building~\location) but not the specific range of values~\notarget (browse). The goal is to identify suitable locations in fa\c{c}ades for urban farming or for green visibility. In this micro-scale analysis, data is uniformly discretized at a building's fa\c{c}ade (see Figure~\ref{fig:scales} (middle, right and bottom, right))~\cite{palliwal_3d_2021, li_room_2022, yu_view-based_2016}.
Also, architects interested in the pedestrian level of comfort leverage wind and sunlight access data at the building and street layers. The analysis involves the comparison of wind comfort maps across multiple months~\cite{de_luca_sun_2019}.

%\noindent \textbf{R1:} \why{Identify} and \why{compare} queries, \what{micro} scale, \what{uniform} discretization, \what{building} and \what{street} physical data entities.

\noindent \textbf{T2. Micro-scale browsing for individual unit identification}

In the property cadastre case, the target data is a volume composed of individual units (e.g., apartments, offices). In other words, individual units assemble into a dense 3D building volume. While in \textbf{T1}, experts were only interested in the outer regions of a building, in property cadastre tasks, architects, land owners, and city officials need to assess attributes from inner units that are potentially occluded.
These attributes include primary function, ownership information, and legal boundaries.
After an individual unit is identified, an expert can further perform shadow or visibility analyses~\cite{shojaei_visualization_2013}, or study the 3D morphology and spatial positions of a building~\cite{ying_distortion_2019}.
% Reducing the occlusion of outer units is then a requirement to analyze the properties of inner units.

%\noindent \textbf{R2:} \why{Identify} query, \what{micro} scale, \what{structural} discretization, \what{building} physical data entities.

\noindent \textbf{T3. Multi-scale exploration for building site identification}

It is common for architects and urban planners to rely on analyses over multiple scales to help identify sites for the development of new buildings.
Differently from \textbf{T1}, which is restricted to a known building, here we have an exploration where both the location \nolocation and identity \notarget are unknown. At each level, the expert will compare distributions leveraging different data.
At a macro scale, an expert usually uses spatial data to compare neighborhoods and better understand strengths and weaknesses between them to establish performance thresholds~\cite{ferreira_urbane_2015}.
At a meso scale, an expert establishes criteria to assess the consequences of new buildings at a neighborhood level, which can include multivariate attributes to measure view impact~\cite{ortner_vis--ware_2017}.
At a micro scale, they can then study the impact of building massing configurations on urban density~\cite{ang_ubemio_2022} and sky exposure at street level.

%\noindent \textbf{R3:} \why{Identify} and \why{compare} queries, \why{distribution} target, \what{micro, meso} scales, \what{structural} discretization, \what{building} and \what{street} physical data entities, \what{multivariate} data.

\noindent \textbf{T4. Multi-scale analysis for scenario comparison}

After identifying a particular site (such as discussed in \textbf{T3}), it is common for experts to have to compare different scenarios for \emph{what-if} analyses. In disaster management, for example, these can be floodwall breach scenarios~\cite{ribicic_visual_2013,cornel_visualization_2015} with data defined at building (e.g., the height of flood) and street (e.g., flooded areas) levels. 
In building energy modeling, the scenarios are different building type parameters (e.g., single-family homes, dense high-rises) that are considered to target certain carbon reduction or energy efficiency goals~\cite{ang_ubemio_2022}.
In view impact and sun access analyses, a scenario is usually a building massing configuration (e.g., number of floors, floor plate size) that impacts the surrounding context~\cite{ferreira_urbane_2015,doraiswamy_topology-based_2015,ortner_vis--ware_2017,miranda_shadow_2019}.
The task then involves analyzing the trade-offs and understanding the potential impact of these interventions (floodwall breach or new buildings) at both micro and meso scales.
Scenario comparison and analyses across scales are also increasingly more common in the urban climate use case~\cite{pacifici_analysis_2019, javanroodi_interactions_2020, jang_decline_2021}, using multi-scale models that couple micro (e.g., numerical models such as CFD) and meso and macro-scale approaches (e.g., WRF, remote sensing maps).

%\noindent \textbf{R4:} \why{Compare} query, \what{micro, meso} scales, \what{building} and \what{street} physical data entities.

\noindent \textbf{T5. Micro and mesoscale analysis of volumetric data}

To better understand the impact of wind on new developments and neighborhoods, experts are increasingly turning to computational fluid dynamic or ENVI-met simulations. These simulations take into account 3D models and dominant wind directions and generate a 2D (at street level) or 3D time-varying flow volume with wind speeds and/or temperature.
The primary task of the expert is to then visualize distributions and (1) identify trends (e.g., increase in wind pressure) at both micro (i.e., individual building facade~\cite{avini_wind_2019}) or meso (i.e., a neighborhood~\cite{hadavi_impacts_2021}) scales, (2) values above certain thresholds~\cite{huang_sensitivity_2023}, and (3) individual features such as jet patterns~\cite{koch_compact_2018}, channeling, sheltering or wake effect~\cite{elshaer_variations_2017}.

%\noindent \textbf{R5:} \why{Identify} query, \why{trend, distribution, features} targets, \what{micro, meso} scales, \what{volumetric, multivariate} data.

\noindent \textbf{T6. Analysis of uncertainties}

We have also found that the analysis of uncertainties is common across different use cases, particularly the ones leveraging data from simulations or models.
For example, when considering flooding simulations, multiple scenarios need to be considered and evaluated for effective disaster prevention~\cite{ribicic_visual_2013}.
In noise mapping, given the complex and dynamic nature of the urban environment, predictions can suffer from underestimation~\cite{lee_assessment_2014}.
The primary task of an expert is to then explore this uncertain data at the macro (e.g., to identify the noisiest areas at building facades or streets) or meso (e.g., to identify effective interventions during disaster events) scales.

\begin{figure*}[ht!]
\begin{center}
\includegraphics[width=1\linewidth]{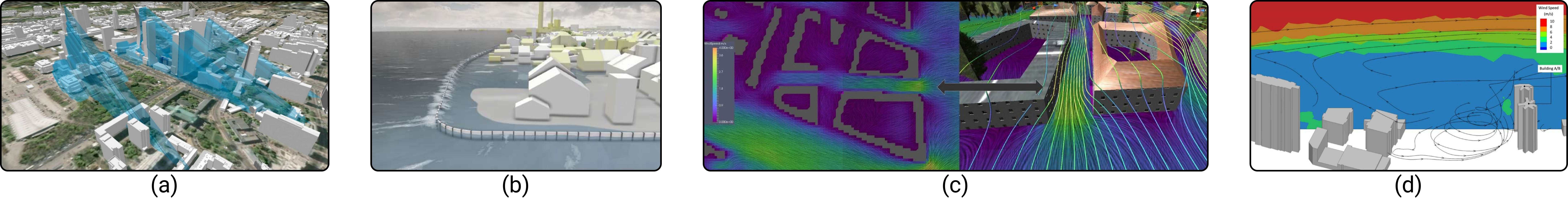}
\end{center}
% \vspace{-0.5cm}
\caption{Different approaches to visualize different urban data defined inside 3D volumetric regions. (a) Shadow volumes~\cite{fijalkowska_assessment_2022}. (b) Realistic water rendering~\cite{cornel_interactive_2019}. (c) Line integral convolution and 3D streamlines~\cite{koch_compact_2018}. (d) Arrow glyphs~\cite{zhang_cfd_2021}.}
\label{fig:3d}
% \vspace{-0.5cm}
\end{figure*}

\section{Visualization \& interaction dimensions (\how{How})}
\label{sec:how}

A fundamental goal of urban visualization is to allow users to investigate the physical properties of a city and relate them to thematic data and vice versa. 
In the case of 3D urban visualization, these physical properties are 3D representations of buildings, streets, or trees, and thematic data, that usually describes complex three-dimensional phenomena, such as wind, microclimate, shadow, visibility, or flooding. 

The design of an effective 3D urban visualization depends on how the thematic data is encoded (Section~\ref{sec:how-encoding}) and by which means they are visually integrated with respective physical data (Section~\ref{sec:how-integration}). 
Figure~\ref{fig:how-what} summarizes the distribution of \how{How} and \what{What} tags.
The extent and verticality of modern cities lead to common challenges such as occlusion and higher cognitive load associated with frequent viewpoint changes. 
To enable users to perform the tasks described in Section~\ref{sec:why}, an effective design needs to deal with occlusions (Section~\ref{sec:how-occlusion}) and support users in navigating urban space (Section~\ref{sec:how-navigation}). Further, we look at how visualization and computational methods are integrated (Section~\ref{sec:how-model}) and which display modalities are used (Section~\ref{sec:how-technology}). Then we conclude this section on how authors typically evaluate their respective contributions (Section~\ref{sec:how-evaluation}).
In this section, we will also highlight \how{How} visualization is used to execute the previously identified tasks.

% by investigating 

%We discuss the visualization of 3D urban data with respect to six dimensions. We classified the reviewed articles based on: How are physical and thematic data visually integrated (Section~\ref{sec:how-integration}); How is thematic data encoded (Section~\ref{sec:how-encoding}); How can a user navigate the 3D scene (Section~\ref{sec:how-navigation}); Which measures are employed to handle occlusions (Section~\ref{sec:how-occlusion}); Which display modality is used (Section~\ref{sec:how-technology}); and how are the respective contributions evaluated (Section~\ref{sec:how-evaluation}).

%%%%%%%%%%%%%%%%%%%%%%%%%%%%%%%%%%%%%%%%%%%%%%%%%%
\subsection{Visual encodings}
\label{sec:how-encoding}
%%%%%%%%%%%%%%%%%%%%%%%%%%%%%%%%%%%%%%%%%%%%%%%%%%

% \nivan{this section seems to follow a different pattern than the others in the sense that it does not include the percentages of the surveyed tag with each tag.}

% \begin{wrapfigure}{r}{0.35\textwidth}
% \begin{center}
% \includesvg[inkscapelatex=false, width=\linewidth]{figs/encoding.vl.svg}
% \end{center}
% \end{wrapfigure}

The integration of physical and thematic data visualizations and the way thematic dimensions are encoded are closely related. We chose to first give an overview of the encodings we have encountered and then support them with examples to facilitate the discussion of their interplay in Section~\ref{sec:how-integration}. In the following two sections, we distinguish between spatial encodings where data is visualized directly in the 3D urban scene, and non-spatial encodings where data is visualized in the form of 2D graphs.

\subsubsection{Spatial encodings}
We distinguish two cases spatially encoding data in a 3D urban visualization. The first is mapping, where thematic data has been mapped and visualized onto physical data and the second is 3D representations to visualize the volumetric context of urban environments.

\noindent \textbf{Mapping onto surfaces.} When mapping data onto buildings, we have found different granularities related to the properties of thematic data discussed in Section~\ref{sec:what-thematic-properties}. 
Nouvel et al. color entire buildings using a discretized color ramp to illustrate their heating demand~\cite{nouvel_influence_2017}. Mapping color in this granularity is also typical for encoding flooding damage to buildings on a neighborhood scale~\cite{waser_many_2014,ribicic_visual_2013, cornel_interactive_2019, yao_3dcitydb_2018}.
In the work of Vanegas et al., buildings are colored depending on their sunlight exposure, the distance to the next park, or their floor-to-area ratio~\cite{vanegas_inverse_2012}. 
Tang et al. present noise-mapping data discretized and visualized per floor level (i.e., discretized over \what{structural} units)~\cite{tang_dynamic_2022}.
Wolosiuk and Mahdavi~\cite{wolosiuk_application_2020} and Li et al.~\cite{li_room_2022} use a \what{uniform} fine-grained mapping to show detailed incident solar radiation on a roof and office view quality on the fa\c{c}ade of high-rise buildings (Figure~\ref{fig:scales} (bottom,right)), respectively. The work by Lorenz et al.~\cite{lorenz_3d_2010} discusses a graphics approach to map 2D rasters and 3D geometries.
Such visualizations directly support the micro-scale fa\c{c}ade-level analyses in \textbf{T1}.

Phenomena that affect both buildings and street and ground levels can be encoded using a \how{3D map} that covers all physical data, for instance, for light~\cite{munoz-pandiella_real-time_2017}, shade~\cite{chow_gis_2014, miranda_shadow_2019}, wind~\cite{avini_wind_2019}, noise~\cite{zhang_case_2017}, and radio propagation~\cite{bounceur_cupcarbon_2018}.
Thematic data encoded in \how{2D map} layers can also be combined with the 3D urban model, for instance, showing the density of subway stations~\cite{ferreira_urbane_2015} or wind force at pedestrian level~\cite{yao_3dcitydb_2018}. However, in these cases, we have found that a fine-grained color mapping onto building fa\c{c}ades is omitted.
Given the nature of the data considered in this survey, it is no surprise that 3D maps are common across all use cases. \highlight{This is largely due to the inherent capacity of 3D visualizations to provide a richer representation and better contextualization of spatial relationships and depth than their 2D counterparts.}
2D maps are frequently utilized in wind and urban climate studies to display data slices at pedestrian height ($\sim$1.75 meter). These maps help visualize and analyze information related to airflow patterns, temperature distribution, and other environmental factors crucial in these cases.

\noindent \textbf{3D representations.}  Phenomena that are defined inside 3D volumetric regions (see Section ~\ref{sec:what-thematic-properties}), for instance, shade, water, or wind, are often encoded into 3D volumetric representations. Fija\l{}kowska et al. show shadow volumes to illustrate when and how many photovoltaic panels are in the shade~\cite{fijalkowska_assessment_2022}. Cornel et al. present a technique to reconstruct a water surface from large-scale shallow water simulation~\cite{cornel_interactive_2019}. They also enhance the visualization of the water by waves and foam to indicate flow direction and barrier overtopping, respectively. 
Beran et al.~\cite{beran_third_2022} propose a set of methods for space-time cube visualization of noise data.
Analysis of wind effects is typically preceded by a CFD simulation.
There is a large design space visualizing such data~\cite{fuchs_visualization_2009}, but most works choose to use \how{glyphs / streamlines} in this context and use the color channel to encode wind speed~\cite{wilkinson_approximating_2014, elshaer_variations_2017, avini_wind_2019, hadavi_quantifying_2020, zhang_cfd_2021, karimimoshaver_effect_2021,athamena_microclimatic_2022} or temperature \cite{hadavi_impacts_2021}.
Koch et al. use illuminated streamlines accompanied by a 2D map using line integral convolution to show wind force on the pedestrian level~\cite{koch_compact_2018}. Another visualization pairs 3D streamlines showing the wind force with a 2D map encoding the ground temperature, similar to Allegrini and Carmeliet~\cite{allegrini_simulations_2018}. 
Zhang et al.~\cite{zhang_cfd_2021}, and Hadavi and Pasdarshahri~\cite{hadavi_impacts_2021} additionally use arrow glyphs on slices of color-coded 2D maps to indicate wind direction.
\highlight{Figure~\ref{fig:3d} highlights some of the approaches.}

It has to be noted that despite the fact that the visualization of CFD data is a core competence of the scientific visualization community, we have not found any systems or design studies dealing with wind analysis in an urban context.
Gaultier et al. use three different geometric representations -- vertical planes, horizontal planes, and points -- to visualize temperature data together with building geometry~\cite{gautier_co-visualization_2020}.
To visualize the seismic response of buildings in an earthquake scenario, Xiong et al. deform the 3D building geometry and color-code the displacement~\cite{xiong_building_2015}. Chen et al. place a ladder truck in the urban scene and illustrate how specific windows can be reached supporting a fire-fighting scenario~\cite{chen_exploring_2017}.
These visualizations support the \textbf{T3} previously identified.

\begin{figure*}[ht!]
\begin{center}
\includegraphics[width=1\linewidth]{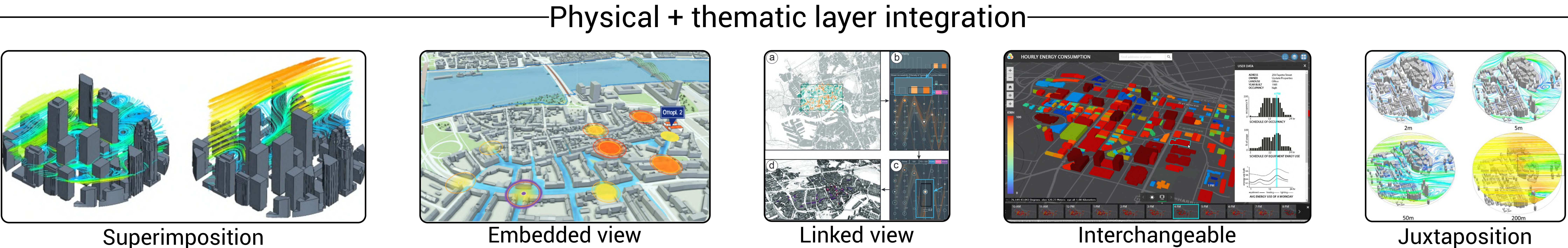}
\end{center}
% \vspace{-0.5cm}
\caption{Examples of approaches to integrate physical and thematic data: superimposition~\cite{elshaer_variations_2017}, embedded views~\cite{cornel_visualization_2015}, linked views~\cite{zeng_vitalvizor_2018}, interchangeable~\cite{krietemeyer_method_2019}, and juxtaposition~\cite{zhang_cfd_2021}.}
\label{fig:physicalthematic}
% \vspace{-0.5cm}
\end{figure*}

\subsubsection{Non-spatial encodings}
%Instead of visualizing thematic data with respect to their given spatialization it can be encoded through chosen spatialization in the form of non-spatial visualizations. 
We have found a diverse set of non-spatial (i.e., abstract~\cite{sorger_taxonomy_2015}) visualizations in the surveyed articles. We also included static visualizations produced in papers that were classified under the \paper{application study} type.
These are not interactive and rarely integrate with spatial views, though we believe they can inform effective visual designs as domain experts are already familiar with these encodings.
Scatter plots, bar charts, and line graphs are widespread among visualization and domain articles across all use cases. 
For example, Miranda et al. use \how{linegraphs} to illustrate changes in light and shadow scores over the course of a year~\cite{miranda_shadow_2019}.
Ribicic et al. visualize flood simulation outputs in \how{bar chart} histograms and \how{scatterplots} to filter interesting runs~\cite{ribicic_visual_2013}.
Krietemeyer et al. encode ``occupancy schedule,'' ``equipment energy use,'' and ``average energy use of a workday'' into two bar charts and a line chart, respectively, as details-on-demand for a building~\cite{krietemeyer_method_2019}.
In the case of wind analysis, we observed frequent use of stacked radial histograms, also referred to as rose diagrams, encoding the frequency of wind of a certain strength and direction \cite{gros_cool_2014, zabarjad_shiraz_wind_2020, avini_wind_2019}.
Fija\l{}kowska et al. employ rose diagrams to show the daily shade and wattage of solar panels~\cite{fijalkowska_assessment_2022}. Hu et al. use polar plots where a 2D scalar field is plotted with respect to zenith and azimuth angles, basically describing a near-hemisphere~\cite{hu_analysis_2019}. 
Heatmaps are employed to display temporal patterns \cite{ouria_evaluation_2018, hu_analysis_2019}, but their axes may also encode positions on a building fa\c{c}ade to show good spots for urban farming \cite{palliwal_3d_2021}, radiation variation~\cite{catita_extending_2014} or show which windows have the best views \cite{li_room_2022}.

When it comes to encoding multiple thematic dimensions, the \how{parallel coordinates} plot (PCP) is a popular choice in the visualization community. 
The PCP can show the plethora of thematic values of a building with relation to all other buildings~\cite{ferreira_urbane_2015, zeng_vitalvizor_2018} or neighborhoods~\cite{butkiewicz_multi-focused_2008,chang_legible_2007} and allows users to assess the performance of building candidates~\cite{zhang_urbanvr_2021} and designs~\cite{doraiswamy_topology-based_2015}. 
PCPs are also used in disaster management to compare the performance of barriers~\cite{ribicic_visual_2013} and rescue paths~\cite{bock_visualization-based_2017}.
PCPs are usually employed in multi-scale analyses (\textbf{T3} and \textbf{T4}) as a way to filter geographical areas that satisfy certain constraints.

There is no PCP in domain contributions except the work of Wilson et al.~\cite{wilson_how_2019} and Su et al.~\cite{sun_generative_2023}, where they combine design parameters and urban indicators of procedurally generated block layouts. 
This might be owed to the unfamiliarity by domain experts but also due to the fact that, in order to be effective, a PCP needs to support interactivity, such as linking \& brushing.
Analysis articles rather use a scatterplot \how{matrix} (SPLOMs) or a correlation matrix to show multiple dimensions at once, for instance, to illustrate correlations between building characteristics and performance indicators \cite{zhang_case_2017, torabi_moghadam_gis-statistical_2018,bonczak_large-scale_2019}.
Ortner et al. use an interactive matrix of candidate buildings and viewpoints where each cell encodes four different view impact metrics as bar charts~\cite{ortner_vis--ware_2017}.
Bock et al. use an interactive SPLOM in addition to their PCP~\cite{bock_visualization-based_2017}.
In order to support \textbf{T6} and uncertainty analyses, domain papers usually employ well-known visualization methods in a static fashion, such as box plots~\cite{lee_assessment_2014}.

%%%%%%%%%%%%%%%%%%%%%%%%%%%%%%%%%%%%%%%%%%%%%%%%%%
\subsection{Integration of physical \& thematic visualizations}
\label{sec:how-integration}
%%%%%%%%%%%%%%%%%%%%%%%%%%%%%%%%%%%%%%%%%%%%%%%%%%

% \begin{wrapfigure}{r}{0.3\textwidth}
% \begin{center}
% \includesvg[inkscapelatex=false, width=\linewidth]{figs/physical-thematic-integration.vl.svg}
% \end{center}
% \end{wrapfigure}

To structure this section, we expand on the related work categorization presented by Mota et al.~\cite{mota_comparison_2022} and derive the following ways of visually integrating physical and thematic data: superimposition, embedded views, linked views, interchangeable, and juxtaposition. Figure~\ref{fig:physicalthematic} shows examples of works using these approaches.

\how{Superimposition} displays two or more data instances in the same coordinate space.
This comes down to the 2D or 3D mapping of thematic data onto physical entities as described in the previous section.
Also, the integration of 3D streamlines, noise, or sunlight access data falls into this category (\textbf{T5}), which is the dominant way of combining thematic and physical urban aspects in the surveyed literature.
A variation of this concept involves streamlining slices with different heights~\cite{zhang_cfd_2021} and orientations~\cite{elshaer_variations_2017} (Figure~\ref{fig:physicalthematic} (Superimposition)). Hadavi and Pasdarshahri~\cite{hadavi_impacts_2021} color-map wind speeds on a semi-transparent vertical plane in addition to a colored 2D layer with arrow glyphs at pedestrian level.
For noise and sunlight access data visualization, the majority of works overlay the data on the surface of buildings~\cite{ranjbar_3d_2012,miranda_shadow_2019, li_noise_2021, wickramathilaka_three_2023}; a notable exception is Beran et al. that propose the use of spacetime cube isopleths~\cite{beran_third_2022}.

%Previously discussed glyphs \cite{cornel_visualization_2015, ortner_vis--ware_2017} are a way of superimposing thematic information as well.

\begin{figure}[t!]
\begin{center}
\includegraphics[width=1\linewidth]{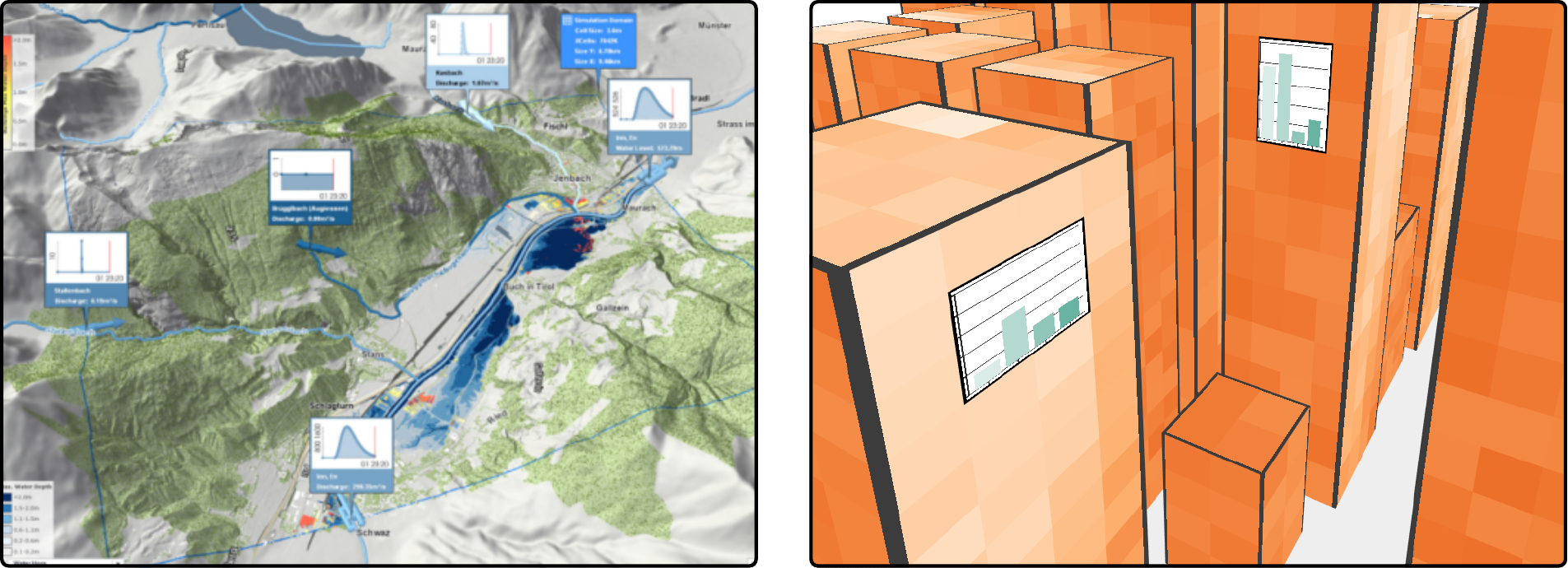}
\caption{\highlight{Examples of embedded views used at two distinct scales. Left: Vuckovic et al.~\cite{vuckovic_combining_2021} showing a wider spatial extent. Right: Mota et al.~\cite{mota_comparison_2022} depicting a local spatial context.}}
\label{fig:embedded}
\end{center}
% \vspace{-0.75cm}
\end{figure}

In \how{embedded views}, views establish their own coordinate system but are spatially anchored in the physical domain~\cite{mota_comparison_2022}.
\highlight{These views combine 2D and 3D graphical elements, in the form of charts and 3D maps~\cite{dubel_2d_2014}.}
%(Figure~\ref{fig:physicalthematic} (Embedded view) and Figure~\ref{fig:embedded}).
%
\highlight{For example, Cornel et al. use circular glyphs embedded on the 3D environment to display flooding uncertainty~\cite{cornel_visualization_2015} as shown in Figure~\ref{fig:physicalthematic} (Embedded view).}
%
% Cornel et al. use so-called fa\c{c}ade area plots that draw water levels of different flooding scenarios directly onto the building fa\c{c}ade~\cite{cornel_visualization_2015}.
%
Butkiewicz et al. anchor histograms and PCPs with a probe metaphor to neighborhoods displaying respective demographic data~\cite{butkiewicz_multi-focused_2008}.
A similar approach is used by Vuckovic et al. to place small hydrographs~\cite{vuckovic_combining_2021} (Figure~\ref{fig:embedded} (left)) and by Wilson et al. to anchor performance indicators in the form of bar charts to urban layouts~\cite{wilson_how_2019}. 
% Cornel et al. \cite{cornel_visualization_2015} place centered bar plots in the 3D scene in a flooding scenario that encode vulnerability per width for each position of a barrier. 
Ortner et al. use color-coded circular glyphs to mark viewpoints in the scene that are impacted by candidate buildings~\cite{ortner_vis--ware_2017}. 
\highlight{Mota et al. evaluate the efficacy of visualizations that are embedded within the physical domain, such as the bar charts shown in Figure~\ref{fig:embedded} (right).}
In all cases, embedded views show detailed thematic information that is confined to a certain region, block, or building.
In our survey, visualization papers were the only examples of the use of embedded views.

Orthogonal to the concept of embedding views is the coordination and \how{linking of views} (Figure~\ref{fig:physicalthematic} (Linked view)).
For instance, selecting a water line in the fa\c{c}ade area plot loads the respective flooding scenario that led to that line~\cite{cornel_visualization_2015}.
However, to avoid clutter and occlusion, non-spatial visualizations are often juxtaposed with the spatial view.
Then linked highlighting or filtering is a way to anchor the encoded thematic data in the physical domain and vice versa. 
A majority of the interactive systems from the visualization community allow linked interaction to highlight and/or filter data.
In contrast to embedded views, non-spatial views in the reviewed cases typically display an overview of all entities instead of a regional subset. Examples are all solutions featuring a PCP~\cite{ribicic_visual_2013, ferreira_urbane_2015, doraiswamy_topology-based_2015, bock_visualization-based_2017, zeng_vitalvizor_2018, zhang_urbanvr_2021}. 
In the case of Krietemeyer et al., the non-spatial views display detailed information about a single building, and the linked highlighting is realized by brushing the temporal domain\cite{krietemeyer_method_2019}.
Linked views have been extensively used to support multi-scale analyses (\textbf{T3} and \textbf{T4}), given that different scales might require the combination of multiple types of encodings.

\begin{figure}[t!]
\begin{center}
\includegraphics[width=1\linewidth]{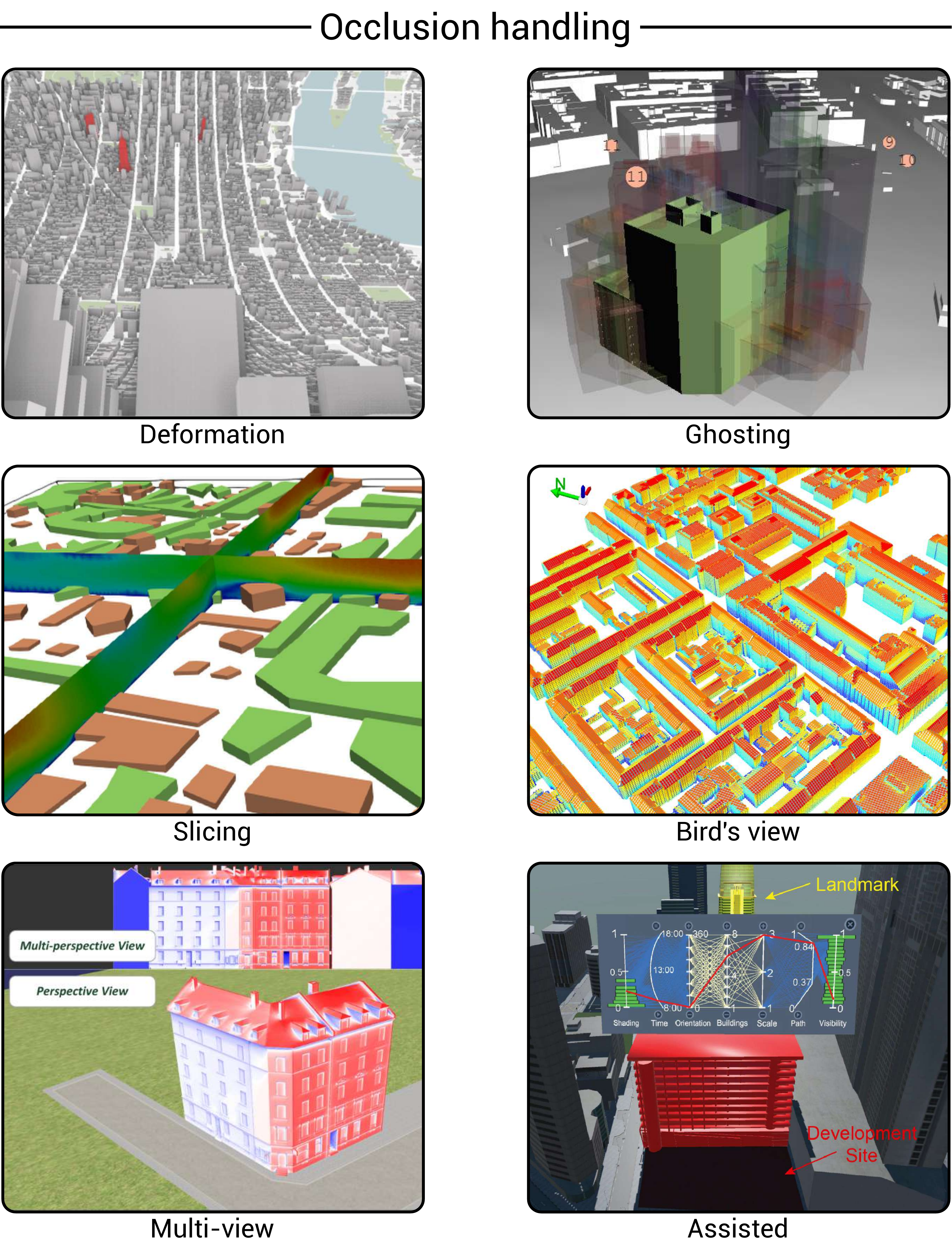}
\end{center}
% \vspace{-0.5cm}
\caption{Examples of approaches to handle occlusion: Deformation~\cite{chen_urbanrama_2022}, ghosting~\cite{ortner_vis--ware_2017}, slicing~\cite{kadaverugu_improving_2021}, bird's view~\cite{willenborg_applications_2017}, multi-view~\cite{pasewaldt_multi-perspective_2013}, assisted-steering~\cite{zhang_urbanvr_2021}.}
\label{fig:oclusion}
% \vspace{-0.5cm}
\end{figure}

For \how{interchangeable} integration, the same view is used to show different states requiring a dynamic visualization.
In Vis-A-Ware, users can flip through building candidates or show them all at once (Figure~\ref{fig:oclusion} (Ghosting))~\cite{ortner_vis--ware_2017}. 
Krietemayer et al. let users browse through different time steps via a filmstrip metaphor (Figure~\ref{fig:physicalthematic} (Interchangeable)) and also provide a slider-based interaction to illustrate before and after scenarios~\cite{krietemeyer_method_2019}. 
With the Shadow Profiler, users can step through or animate the course of shadows cast on an urban scene~\cite{miranda_shadow_2019}.

Authors often \how{juxtapose} spatial views when different states of the physical domain shall be compared. 
This works especially well when the spatial domain can be reduced to a region of interest such as a city block~\cite{wilson_how_2019} or a single building~\cite{doraiswamy_topology-based_2015}.
In the domain of urban climate and wind analysis, we have found several application studies that convey their results by juxtaposing static visualizations. 
Wilkinson et al. illustrate the difference between a ground truth simulation and their approximate wind interference model by two top-down views showing streamlines and three views of a color-mapped building, where the third view encodes the difference~\cite{wilkinson_approximating_2014}. 
Elshaer et al. extensively use this concept, most notably circular cut-outs that show a vertical and a horizontal streamline slice for three different time steps~\cite{elshaer_variations_2017}, similar to Zhang et al. who show circular urban cutouts overlaid with horizontal streamline slices at different heights (Figure~\ref{fig:physicalthematic} (Juxtaposition))~\cite{zhang_cfd_2021}. 
Hadavi and Pasdarshahri juxtapose four views showing two different buildings overlaid with detailed 3D streamlines with color-coded velocity in the upper row and color-coded temperature below~\cite{hadavi_impacts_2021}.
Pasewaldt et al. juxtapose the 3D view (perspective view) of a building and a 2D view (multi-perspective view) of its unwrapped fa\c{c}ade, both mapped with thematic data from solar potential analysis~\cite{pasewaldt_multi-perspective_2013}.
Juxtaposed spatial views have been found to be particularly popular in papers where the primary task is the comparison of multiple scenarios (\textbf{T4})~\cite{doraiswamy_topology-based_2015, ortner_vis--ware_2017}.

In our review, we have found that ArcGIS~\cite{johnston_arcgis_2001}, UrbanSim~\cite{noth_urbansim_2003} and QGIS~\cite{qgis_2023} support superimposition, linked view, interchangeable and juxtaposition.
Kepler.gl~\cite{kepler_2023} and deck.gl~\cite{wang_deckgl_2017} support superimposition, linked view, and interchangeability but have limited support for juxtaposition.
None of these tools and libraries provide support for the creation of embedded views in the 3D environment.

%The vast majority of works rely on `overlay' strategies (also called `superimposition' \cite{kim_comparison_2017} or `superposition' \cite{gleicher_considerations_2018}) by visualizing the thematic data in the physical coordinate space.

%\textit{Points to mention: (1) Most works rely on overlay strategies. (2) Works that use linked view are fully-fledged systems covering multiple resolutions.}

%%%%%%%%%%%%%%%%%%%%%%%%%%%%%%%%%%%%%%%%%%%%%%%%%%
\subsection{Occlusion handling}
\label{sec:how-occlusion}
%%%%%%%%%%%%%%%%%%%%%%%%%%%%%%%%%%%%%%%%%%%%%%%%%%

% \fabio{Add slicing, mention cross cuts and side views: \cite{javanroodi_interactions_2020,shareef_effect_2020,shaeri_investigation_2018,beran_third_2022}}

% \begin{wrapfigure}{r}{0.35\textwidth}
% \begin{center}
% \includesvg[inkscapelatex=false, width=\linewidth]{figs/occlusion.vl.svg}
% \end{center}
% \end{wrapfigure}

%Points to mention: (1) not many works on assisted steering. (2) bird's eye view popularity.

The verticality and density of cities easily lead to the occlusion of analysis targets. As a result, the analysis tasks described in Section~\ref{sec:why} are impeded or may fail. 
We classify the occlusion handling measures of the surveyed articles based on six categories: deformation, ghosting, multi-view, assisted steering, slicing, and bird's view.
This is an adaptation of the five occlusion handling design patterns introduced by Elmqvist and Tsigas~\cite{elmqvist_taxonomy_2008}. 
Most notably, we changed ``tour planner" to assisted steering, which conveys the more interactive nature of the found techniques, and we added bird's view, which is a ubiquitous remedy for occlusion problems in an urban context. We show examples for each technique in Figure~\ref{fig:oclusion}.

% \thomas{we start here deformation while figure 14 starts with ghosting. slicing seems new to me - is this an adaptation of Elmqvist and Tsigas volumetric probe?, maybe needs introduction - we also would need an example image}

Several \how{deformation} techniques alter the 3D urban model primarily by scaling to disocclude routes and landmarks~\cite{huamin_qu_focuscontext_2009, hirono_constrained_2013}.
In addition to disocclusion through scaling, Deng et al. also guide users to a bird's eye perspective and demonstrate the potential of their system by encoding a 3D pollution layer~\cite{deng_interactive_2016}.
Wu and Popescu render multiple view angles from a single position and compose them to a distorted view of an urban scene~\cite{wu_multiperspective_2016}. 
Chen et al. present a method for dynamically ``exploding'' urban geometry to reveal a point of interest on the ground in an AR setting~\cite{chen_immersive_2017}.
Ying et al. propose a similar approach for the identification of individual units in property cadastre (\textbf{T2})~\cite{ying_distortion_2019}.
To increase viewing distance in a CAVE setup Engel et al. bend a sparse city scene upwards~\cite{engel_immersive_2012}.
In a fully immersive VR setting, Chen et al. wrap the dense cityscape of Manhattan upwards onto a cylindrical projection to keep distant landmarks in sight (Figure~\ref{fig:oclusion} (Deformation))~\cite{chen_immersive_2017}.

Cornel et al. let users focus on a target building and cut away all occluding geometry while using \how{ghosting} for the outlines of adjacent buildings for context~\cite{cornel_visualization_2015}. The system presented by Lv et al. enables users to inspect individual floor layouts. When a floor is selected, the floors above it are cut away~\cite{lv_managing_2016}.
Seipel et al. investigate how different transparency settings influence a person's perception of complex 3D real property cadastre data~\cite{seipel_visualization_2020}.
Gaultier et al. use ghosting on thematic data encoded by 3D glyphs to blend temperatures with the urban geometry~\cite{gautier_co-visualization_2020}, supporting \textbf{T5}.
Robineau et al. adjust the transparency of built environment features to highlight wind streamlines~\cite{robineau_coupling_2022}.

Pasewaldt et al. take a \how{multi-view} approach where the 3D view of a building is supported by a multi-perspective view that unwraps the fa\c{c}ade onto a 2D plane effectively removing occlusion (Figure \ref{fig:oclusion} (Multi-view))~\cite{pasewaldt_multi-perspective_2013}. 
A similar approach has been employed to visualize the distribution of view impact~\cite{salimi_visual_2023}, temperature~\cite{haeri_evaluation_2023}, and sunlight access~\cite{catita_extending_2014,palliwal_3d_2021} over building fa\c{c}ades.
Zhang et al. support user navigation with \how{assisted}-steering by automatically computing optimal viewpoints which try to minimize occlusions (Figure~\ref{fig:oclusion} (Assisted))~\cite{zhang_urbanvr_2021}. Deng et al.~\cite{deng_interactive_2016} and Li et al.~\cite{li_seevis_2020} were the only other works that made use of assisted steering.

Another popular approach, particularly for volumetric data in wind, urban climate, and noise use cases, is the use of \how{slice} planes. The concept is similar to volumetric probes, in which the occlusion of targets is handled through the use of user-defined cutting planes (Figure~\ref{fig:oclusion} (slicing)).
In the surveyed works, we have found two common slice-based approaches. In the first group, we have works that make use of \emph{static} horizontal and vertical cross sections~\cite{shaeri_investigation_2018,javanroodi_interactions_2020,shareef_effect_2020}. For example, Shaeri et al. use vertical cross sections at different sections of a neighborhood and horizontal cross sections at different heights for the analysis of air speed and air pressure~\cite{shaeri_investigation_2018}.
Another approach is to use cross-sections embedded in an interactive 3D environment. For example, Kadaverugu et al. superimposes vertical cross-sections within a 3D urban environment to slice a CFD domain~\cite{kadaverugu_improving_2021}, and Ranjbar et al. and Beran et al. use cross-sections for 3D noise visualization~\cite{ranjbar_3d_2012, beran_third_2022}.

Occlusion handling is a common topic in 3D visualization research~\cite{elmqvist_taxonomy_2008}, and several contributions exist in the context of 3D urban scenes. However, they barely find their way into 3D urban analytics solutions that feature rich thematic data, as they are present in most use cases. The majority of spatial visualizations rely on the users taking a \how{bird's view} through manual camera manipulation. Naturally, this is the preferred method for non-interactive visualizations, such as in Willenborg et al. (Figure~\ref{fig:oclusion} (Bird's view))~\cite{willenborg_applications_2017} and most other application studies.

%%%%%%%%%%%%%%%%%%%%%%%%%%%%%%%%%%%%%%%%%%%%%%%%%%
\subsection{Navigation methods}
\label{sec:how-navigation}
%%%%%%%%%%%%%%%%%%%%%%%%%%%%%%%%%%%%%%%%%%%%%%%%%%

% \begin{wrapfigure}{r}{0.3\textwidth}
% \begin{center}
% \includesvg[inkscapelatex=false, width=\linewidth]{figs/manipulation.vl.svg}
% \end{center}
% \end{wrapfigure}

%Points to mention: (1) most works just use manipulation-based navigation. (2) Lack of steering mechanisms (drill down to use cases and resolutions to see the distribution across them).

Navigation poses one of the main challenges when visualizing 3D urban environments. Observing physical and thematic data at different scales and heights impeded by occlusion requires frequent viewpoint changes that lead to a loss of context and can put a high cognitive strain on the users. In accordance with the travel metaphors presented by LaViola et al.~\cite{laviola_interfaces_2017} we distinguish walking, steering, selection-based travel, and manipulation-based travel in the reviewed articles.
%We distinguish four different forms of navigation walking, steering, selection-based, and manipulation-based, in accordance with LaViola et al.~\cite{laviola_interfaces_2017} \nivan{which ones}?. 

UrbanRama is a fully immersive urban visualization where the environment is wrapped upwards onto a cylindrical projection mixing local and global views~\cite{chen_urbanrama_2022}.
Beyond \how{walking} and fly-based \how{steering}, also present in other immersive solutions~\cite{engel_immersive_2012, bartosh_immersive_2019}, they offer a sling-shot transition after users select a building in the distance or on a map metaphor.
Systems dealing with view analysis typically provide \how{selection}-based navigation to put users into the perspective of a specific view point~\cite{doraiswamy_topology-based_2015, ferreira_urbane_2015, ortner_vis--ware_2017}. 
Pasewaldt et al. juxtapose an unwrapped multi-perspective view and a 3D perspective view of a building to support users with bi-directional selection-based navigation and view manipulation~\cite{pasewaldt_multi-perspective_2013}.
Users can select a point on the 3D fa\c{c}ade, and the view onto the 2D surface will zoom in on that region.
The selection of points on the unwrapped 2D surface will trigger respective camera transitions in the perspective view.
The vast majority of contributions that feature some 3D visualization of the urban environment, be it a GIS-like application or a simulation tool, make use of \how{manipulation}-based navigation where users manually change the view position and orientation.
Doing so without any guidance approaches potentially results in the aforementioned context loss and high cognitive load, particularly impacting tasks that straddle multiple scales (\textbf{T3} and \textbf{T4}).

%%%%%%%%%%%%%%%%%%%%%%%%%%%%%%%%%%%%%%%%%%
\subsection{Visual analytics systems}
\label{sec:how-model}
%%%%%%%%%%%%%%%%%%%%%%%%%%%%%%%%%%%%%%%%%%

% \begin{wrapfigure}{r}{0.35\textwidth}
% \begin{center}
% \includesvg[inkscapelatex=false, width=\linewidth]{figs/va-integration.vl.svg}
% \end{center}
% \end{wrapfigure}

A fundamental aspect of 3D urban analytics is the integration of computational models with visualization and analysis processes. In their survey, Deng et al. distinguish four levels of integration: visual analytics without models (i.e., urban analyses relying on well-designed data visualizations), post-model visual analytics (models are used to discover knowledge and visualization to present it), model-integrated visual analytics (close integration between models and visualizations), and visual analytics-assisted models (visualizations used to diagnose, adjust or improve models)~\cite{deng_survey_2023}.

\how{Visual analysis without models} is common in 3D property cadastre~\cite{shojaei_visualization_2013,seipel_visualization_2020}, where tasks often require the analysis of property data without a computational model involved.
The \how{post-model visual analytics} tag applies to the vast majority of domain contributions where a simulation or model is run, and then the results are analyzed (application study) or a system is produced that allows users to explore the results. 
Moreover, papers from the visualization community that focus on visualization and interaction contributions often fall into this category. For instance, Vis-A-Ware, where users compare building viewpoint relationships but can neither define new viewpoints nor building candidates~\cite{ortner_vis--ware_2017}, VitalVizor which allows users to explore a city with respect to urban vitality metrics while having no influence on the metrics computation~\cite{zeng_vitalvizor_2018}, and The Urban Toolkit~\cite{moreira_urban_2024}, a grammar-based framework for the construction of urban visual analytics tools.

Considering the \how{model-integrated visual analytics} tag, Wilson et al. present a system that allows users to interact with an engine that creates block layouts and automatically computes performance indicators such as daylight or comfort~\cite{wilson_how_2019}. 
Among visualization contributions, Vanegas et al. support city development with an inverse design approach to adapt city layouts to match constraints, such as lowering the average distance to the nearest park~\cite{vanegas_inverse_2012}. 
Another design example comes from Doraiswamy et al., where users can pick and steer tower designs depending on the resulting view quality and in adherence to floor programs~\cite{doraiswamy_topology-based_2015}. 
With Urbane~\cite{ferreira_urbane_2015} and Shadow~Profiler~\cite{miranda_shadow_2019}, it is possible to interactively evaluate and compare new building developments depending on their view and shadow impact, respectively. 
Zhang et al. address a similar use case by letting users position, rotate, and scale a candidate building while optimal viewpoints for navigation and visibility metrics are calculated. In the domain of flood management, several works feature tight integration with a simulation model~\cite{zhang_urbanvr_2021}. 
Ribicic et al. use linked scatterplots, bar charts, and a PCP to explore and evaluate barrier performance with respect to multiple simulation runs~\cite{ribicic_visual_2013}. 
Barrier plans can be updated and instantly reevaluated. Waser et al. extend this work to scale well with a large number of flooding scenarios and respective prevention plans~\cite{waser_many_2014}.

Works with the \how{visual analytics-assisted models} include Konev et al.\cite{konev_run_2014}. Their system RunWatchers monitors, terminates, and starts new flood simulation runs from existing ones according to domain-specific rules. 
This leads to a more efficient traversal of the search space and improvement of overall runtimes due to the early termination of failed runs.
With the same tag, Lyu et al. recently proposed a visual analytics framework for fair urban planning~\cite{lyu_if_2023}; the framework uses a 3D map to facilitate the understanding of spatial context, including building shapes and heights, as well as to increase immersion at street level.

%Points to mention: (1) Taxonomy from Deng et al's survey. (2) Mention the one VA-assisted modeling. (3) Overwhelming majority of works is post-model VA.

%%%%%%%%%%%%%%%%%%%%%%%%%%%%%%%%%%%%%%%%%%%%%%%%%%
\subsection{Display modalities}
\label{sec:how-technology}
%%%%%%%%%%%%%%%%%%%%%%%%%%%%%%%%%%%%%%%%%%%%%%%%%%

% \begin{wrapfigure}{r}{0.325\textwidth}
% \begin{center}
% \includesvg[inkscapelatex=false, width=\linewidth]{figs/tech.vl.svg}
% \end{center}
% \end{wrapfigure}

%Points: (1) overwhelming majority of desktop use cases. (2) no idea what else to mention, perhaps remove this? 

% The vast majority of surveyed solutions run in \how{desktop} environments. Hence we will take a closer look at alternative display modalities, as there are CAVE, VR, AR, and mobile. 

The vast majority of surveyed solutions run in \how{desktop} environments, with domain-specific works using a combination of desktop-based GIS tools for visualization (e.g.,~ArcGIS, Rhino, SAGA, Cesium ion), backed up by specific libraries (e.g.,~PySolar, LIS~Pro~3D) for analyses and standard databases~(e.g., PostGIS) for data management.
Considering the visualization papers, their contributions rely on custom-built frameworks that leverage low-level libraries (e.g., OpenGL, WebGL) or game engines (e.g., Unity).

We have also found contributions using alternative display modalities.
To convey 3D phenomena or urban layouts, users can be immersed in a virtual cityscape in the form of a CAVE or VR setup.
Engel et al. present a 360-degree urban scene from the pedestrian level in a \how{CAVE} to support collaboration and decision-making when dealing with complex spatial information~\cite{engel_immersive_2012}.
Perhac et al. present a system with VR capabilities for the visualization of social media data~\cite{perhac_urban_2017}, primarily focusing on its use by urban planners.
Bartosh et al. take the concept of a classical GIS system and translate it to \how{VR}~\cite{bartosh_immersive_2019}.
Users can inspect layers encoding thematic information from a literal bird's eye view or inspect buildings from a pedestrian level. 
Building details are displayed with abstract charts in the form of a virtual fact sheet.
Chen~et al. do not investigate thematic data visualization but offer a deformation technique and interaction metaphors to seamlessly explore local and global urban views~\cite{chen_urbanrama_2022}. 
Chen~et al. introduce an exploded view technique for immersive urban environments to deal with occlusion problems \cite{chen_immersive_2017}.
Koch et al. argue that an immersive environment is suitable to illustrate complex wind flow behavior in combination with temperature and to convey the impacts of urban developments to the public~\cite{koch_compact_2018}.
In their visualization, they show 3D color-coded streamlines and a color layer on the ground encoding temperature and velocity. 
Zhang et al. use a VR environment to let users intuitively place and manipulate a candidate building~\cite{zhang_urbanvr_2021}. 
Apart from computed performance indicators, users can also view candidates from a pedestrian level and qualitatively judge their impact on the cityscape. 
In the context of urban search and rescue scenarios, Bock et al. offer incident commanders an immersive view of collapsed buildings to judge the size of rescue paths~\cite{bock_visualization-based_2017}.
We only found two systems for \how{mobile} use. XEarth translates a classical GIS application to a mobile display~\cite{li_xearth_2015}, while Chen et al. provide fire brigades en-route with in-situ information, for instance, where and how to place ladder trucks to access fire sources~\cite{chen_application_2014}. 
The exploded view technique by Chen et al. presents the only contribution using \how{AR} technology~\cite{chen_immersive_2017}. 
The low number of immersive and in-situ contributions may indicate a research gap, however, it is likely that such systems are mostly published in more modality-specific venues (e.g., IEEE VR, ISMAR).

%\thomas{I would say most solutions are desktop based but discuss the few that are not in detail and relate them to usecases}

% \marcos{Should we discuss that 3D urban data analytics still needs do be developed?}

%%%%%%%%%%%%%%%%%%%%%%%%%%%%%%%%%%%%%%%%%%%%%%%%%%
\begin{figure}[t!]
\begin{center}
\includegraphics[width=\linewidth]{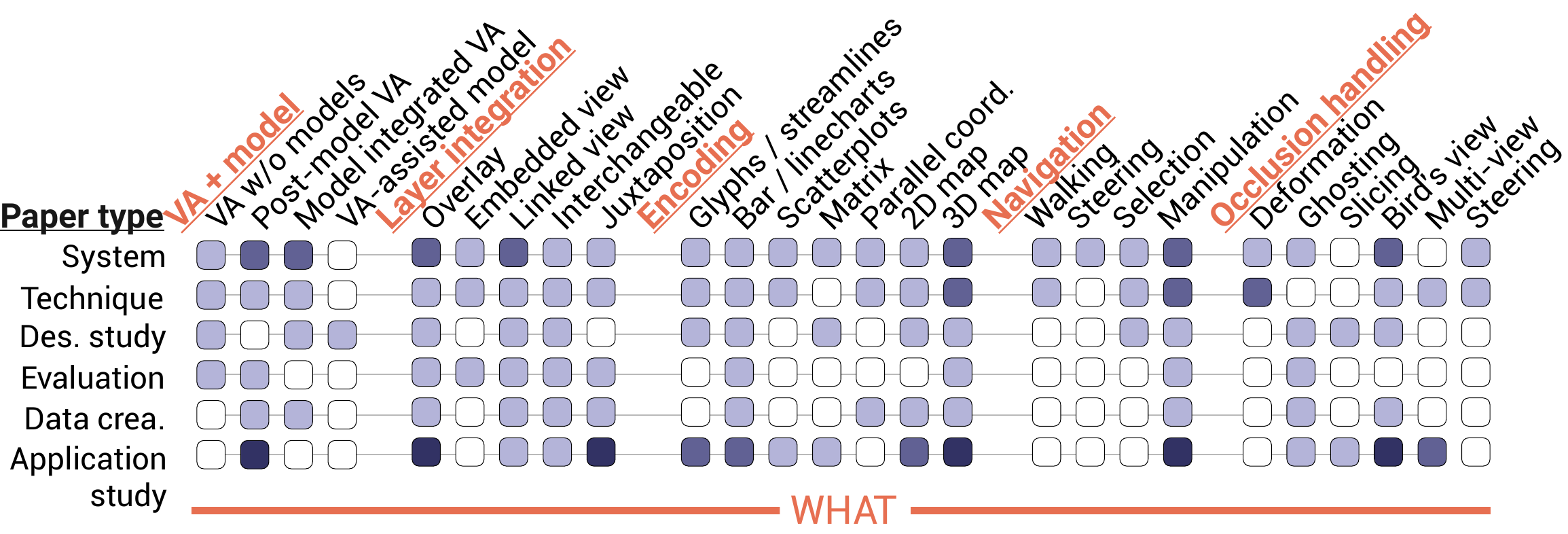}
% \vspace{-0.5cm}
\caption{Distribution of papers according to \paper{Paper type} and \what{What} dimensions, with shades denoting tag occurrence. Data creation and application study refer to domain-specific papers.}
\label{fig:type-how}
\end{center}
% \vspace{-0.75cm}
\end{figure}

\subsection{Evaluation methods}
\label{sec:how-evaluation}
%%%%%%%%%%%%%%%%%%%%%%%%%%%%%%%%%%%%%%%%%%%%%%%%%%

%In our literature review, we coded four different evaluation types: \how{case study}, \how{user study}, \how{statistical evaluation}, and \how{expert interviews}. 
\highlight{In our literature review, we coded the following four evaluation types: \how{Case studies}, where the authors evaluate their contributions by presenting a use case, typically using real data without involving test subjects. Case studies are the dominant evaluation method among domain papers, especially of the domain-specific types \paper{data creation} and \paper{application study}. 
Some of theses works also introduce new methods. In this case, authors typically choose to perform a \how{statistical evaluation} in addition to the case study. Statistical evaluation describes a form of automated evaluation and validation against a baseline or ground truth without involving human test subjects, for instance in the form of validating a CFD simulation against ground truth from wind tunnel experiments~\cite{zhang_cfd_2021}.
%
%Considering the visualization papers, every design or system contribution featured expert interviews often combined with a case study and/or user study.
%

In a \how{user study}, participants are enlisted to execute a sequence of tasks in a controlled testing environment. Results can be quantitative or qualitative, such as measuring time or accuracy regarding tasks or asking their opinion about usability via Likert scale, respectively.}
In the surveyed papers, user studies are often used to evaluate a specific visualization technique~\cite{huamin_qu_focuscontext_2009, hirono_constrained_2013, deng_interactive_2016, chen_urbanrama_2022,beran_third_2022} or system~\cite{bock_visualization-based_2017, zhang_urbanvr_2021}. Herbert and Chen~\cite{herbert_comparison_2015}, Seipel et al.~\cite{seipel_visualization_2020}, and Mota et al.~\cite{mota_comparison_2022} are examples of studies that evaluate aspects of 3D urban visualization (visualization properties, rendering attributes, and embedded plots, respectively).

\highlight{Finally, \how{expert interviews} are evaluations where domain experts (and users of the software) give feedback usually in the form of structured interviews. Compared to user studies, participants are fewer and the feedback is rather interpreted than quantified. Combining a case study with expert interviews and/or a user study is the dominant evaluation pattern for \paper{design study} and \paper{system} papers from visualization venues.}
\paper{System} contributions presented in domain-specific venues rarely use expert feedback with two exceptions~\cite{ang_ubemio_2022, engel_immersive_2012}. Li et al.~\cite{li_xearth_2015}, and Lv et al.~\cite{lv_managing_2016} conduct small-scale user studies to evaluate interactions in their GIS applications.

\section{Discussion, research directions \& opportunities}
\label{sec:new_challenges}

Despite all the surveyed contributions, 3D urban visual analytics is still incipient, especially compared to its 2D counterpart.
In fact, the design of systems for 3D urban data analytics presents a large number of research opportunities in different areas, which we discuss next.
In this section, we discuss a series of takeaways from our survey, focusing on visualization researchers.
To help this discussion, Figure~\ref{fig:type-how} shows the distribution of papers according to paper type.

\subsection{Visual metaphors for thematic dimensions}

While the acquisition technologies and domain analyses are becoming more sophisticated at a fast pace, the visual metaphors used to support visual analysis in 3D environments have not progressed accordingly.
The majority of works rely on superimposition and linked view approaches to visualize 3D urban data. From a technological point of view, these are the most straightforward ways to integrate physical and thematic data. However, embedded views have been shown to improve user performance in certain analytical tasks when considering univariate data~\cite{mota_comparison_2022}.

\noindent \textbf{Common analytical scenarios.} Typical applications in urban planning involve not only studying the current state of a city but also analyzing possible scenarios for development (\emph{scenario planning}) that help business decisions and policy making.
In such situations, thematic dimensions are often used to visually represent the criteria used to assess the quality of the physical elements (present or future) in the city.
Designing visual metaphors that enable the exploration of these thematic dimensions in their spatial context is not a trivial task.
For example, given the complexity of cities, this can result in complex visualization scenarios such as multi-variate~\cite{doraiswamy_topology-based_2015} or temporal analysis~\cite{mota_comparison_2022}, these were common in a large number of surveyed works.
In addition, these thematic dimensions have to be explored in multiple resolutions. For example, domain experts who analyze the quality of views along the surface of a building may be interested in analyzing this quality over a high-resolution grid or by semantic elements like floors, fa\c{c}ades, or even for the entire building.
In such cases, visual representations that depict the distribution of aggregated values are desired~\cite{elmqvist_hierarchical_2010,srabanti_comparative_2022} but have not yet been applied in the 3D urban context.
\highlight{In summary, there is a pressing need for studies that design (possibly extending visual metaphors used in 2D) and evaluate visual metaphors for multi-variate and temporal data in 3D urban environments. This will enable a better support for recurring analytical scenarios in important application domains.}

\noindent \textbf{Uncertainty.} The thematic dimensions used in the analysis of scenario planning are commonly the result of geometric computations or statistical modeling and, therefore, are naturally subjected to inaccuracies and uncertainty.
While being of great importance for spatial analysis in general~\cite{slusarski_visualisation_2019} and urban analysis in specific (\textbf{T6}), almost no attention has been given to the design of new metaphors for the visualization of uncertainty for 3D urban data.
Therefore, there is a need to investigate new designs to visualize uncertainty within an urban environment, as well as a more cohesive integration between urban visualization and works tackling uncertainty in information visualization~\cite{kamal_recent_2021}.
Another important direction is the design of visualizations that allow for the analysis of the interplay between model uncertainty and spatial context, particularly for building energy modeling, in which neighboring buildings play an important role in modeling accuracy and results~\cite{wen_fast_2022}.
There is also a need for work that provides a more grounded understanding of how different urban stakeholders (not only researchers but also policymakers) consider data uncertainties within the urban environment, and, if they do not consider it, understand their reasons~\cite{hullman_why_2020}.
While previous work has explored how multiple data-heavy domains deal with the different aspects of uncertainty~\cite{boukhelifa_how_2017}, no work has characterized how urban visualizations workflows are influenced by uncertainty and, especially, their propagation and impact on decision-making and policy.
This becomes even more pressing given the implications of the data on public policies and community participation~\cite{wan_developing_2019, dembski_digital_2019, hong_ten_2020, botin_digital_2022, wang_data_2022}.

\subsection{Data handling}

\noindent \textbf{Data formats.} In our survey, we have found that there is a great diversity of formats and standards for 3D city geometry and thematic data.
While some local governments make buildings' geometry available (either as meshes or XML-based formats such as CityGML)~\cite{2023ijgis3dcityindex}, studies often rely on crowdsourced data such as OpenStreetMap for a description of the built environment~\cite{2023_bae_osm_qa}.
Such data requires parsers to handle missing data and complex geometries, and translating the data from or to CityGML might lead to inconsistencies.
This multitude of formats and standards leads to ad-hoc data management solutions and hinders the investigation of new visual queries that integrate semantic (i.e., structural) elements of the built environment with thematic data for exploratory analysis.

\noindent \textbf{Data management for interactive visualization.} One important aspect of interactive data analytics is the fact that the latency experience from the interaction to the final visualization is important for the user experience and overall data exploration effectiveness~\cite{liu2014effects}.
For this reason, a large number of works have devoted attention to designing data handling strategies to support interactive exploration of large datasets~\cite{doraiswamy_interactive_2018,battle2021structured}.
In particular, a large body of this work was designed to support the exploration of spatiotemporal data, where the spatial component is 2D.
These strategies support operations such as filtering and hierarchical aggregation of thematic data for large datasets.
However, these solutions have not been extended to 3D spatial data.
For this reason, current 3D urban analytics systems use ad hoc data handling solutions, which often limit their data exploration capabilities and/or their performance.
With the increase in data availability, there is a strong need for solutions that could support interactive explorations of 3D urban data.
In particular, such a solution has to be expressive to allow for queries that align with the semantic elements in a city: buildings, fa\c{c}ades, floors, windows, streets, neighborhoods, etc. 
Also, it has to be efficient to support the interactive navigation, filtering, and change in the aggregation of the data.
Furthermore, this solution needs to allow for the generation and indexing of varied thematic dimensions, which is of great importance for scenario planning~\cite{ferreira_urbane_2015,doraiswamy_topology-based_2015}.

\subsection{Navigation and guided explorations}

\noindent \textbf{Navigation.} As the scale of data grows (larger spatial coverage and/or higher resolution), there is a need for innovative mechanisms to guide data exploration and visualization, with a particular emphasis on accommodating the complexities of 3D interactions.
This is especially true when considering more complex scenarios such as temporal or multi-variate thematic data and highly occluded dense environments (e.g., large urban centers).
In these cases, interesting urban patterns can happen in multiple (spatial/temporal) scales, and the process of searching through it by basic pan, zoom, and filtering can be ineffective.
For navigation, most of the surveyed papers rely on standard 3D manipulations, with no domain-specific paper making use of walking, selection, or steering.
We have also not found any sunlight access paper using walking as navigation, despite shadow playing an important role for the comfort of pedestrians~\cite{Middel2016}.
\highlight{Moreover, surveyed works use standard techniques in which the users have direct control over the navigation process. A promising avenue for future research involves the exploration of sketch-based navigation coupled with analytics within an urban environment. This could involve, for example, investigating of the applicability of Hagedorn and D\"{o}llner's sketch-based navigation~\cite{hagedorn_sketch_2008} or Danyluk et al.'s recent proposal~\cite{danyluk_look_2019} for the analysis of data within 3D urban environments.}

\noindent \textbf{Guided explorations and occlusion handling.}
Methods that help users automatically or semi-automatically select points of view or data slices to inspect, can be effective in helping users to find interesting patterns.
In the case of 2D maps, there have been several techniques that help users to target specific portions of the thematic data, i.e., help to locate interesting points in space and time for further inspection. 
For example, the techniques proposed by Doraiswamy et al.~\cite{doraiswamy_using_2014}, Valdivia et al.~\cite{valdivia_wavelet-based_2015}, and Liu et al.~\cite{liu_tpflow_2019} perform spatiotemporal pattern mining based on computational topology, graph signal processing, and tensor decomposition, respectively. In the context of scientific visualization, multiple works have proposed approaches to facilitate the exploration of volumetric data~\cite{beyer_connectomeexplorer_2013,gu_mining_2016,pandey_cerebrovis_2020}. Ortner et al.~\cite{ortner2016visual} propose a methodological framework on the integration of 3D spatial and non-spatial visualizations. However, they do not offer concrete implementations tailored to 3D urban geometry nor the integration of temporal data. 
Such techniques have not been applied in the context of 3D urban visualization systems. 
One important aspect to notice is that their computational costs and their integration with the other elements in the visual interface of a system present significant challenges.

In the case of occlusion handling, bird's view is still the most popular approach in the surveyed works across all use cases, which poses a problem given the impact that occlusion has in the different tasks highlighted in the survey.
None of the domain papers make use of more complex occlusion handling techniques, such as distortion, multi-view, or steering.
Automated techniques exist to solve the related problems of disocclusion~\cite{deng_interactive_2016}, viewpoint management, and route design~\cite{neuville_3d_2019}.
However, these techniques try to select the best viewpoints of a target object once this object is identified.
For these reasons, these have often been used to only target physical elements of the urban environment, such as landmarks and street paths.
We envision that these could be coupled with techniques for pattern mining, such as the ones cited above, to provide thematic data-based occlusion handling and navigation.
This will help users reduce the burden of excessive interaction to focus on spatial patterns while preserving the perception of the spatial context and facilitating tasks that straddle multiple scopes (\textbf{T3} and \textbf{T4}).

\subsection{Empirical validation of visual designs}

While many 3D urban visual analytics systems have been proposed, few empirical studies were done in these scenarios to evaluate the visualizations, interactions and the overall human experience and exposure within urban spaces (see Section~\ref{sec:how-evaluation}).
This oversight suggests a slight potential disconnect with the latest human-centric developments in digital twins and geospatial science~\cite{Resch2019,2023_scs_human_dt}.
As discussed throughout this survey, the 3D urban environment adds a great deal of complexity to the navigation, interactions, and occlusion handling. 
While some evaluations were done in each of these cases, they are often explored in isolation~\cite{chen_urbanrama_2022}. 
This means that understanding how effective these techniques will be for data analytics is largely an open problem.  
Furthermore, as seen in Section~\ref{sec:paper_types}, the vast majority of the works concentrate on the development of the systems and on the application domain and not on the evaluation of the visual designs used.
This is especially critical when considering the inclusivity and accessibility requirements of various user groups. Creating inclusive visual analytics systems calls for recognizing and addressing the unique challenges faced by individuals with varying abilities (e.g., visual impairments) that may result in considerations of alternative modes of presentation, visualization, navigation, and interaction.
As a result of this, the visualization metaphors and interactions used are, in many cases, done without clear guidance~\cite{neuville_3d_2019} or, at best, based on principles and studies done in 2D scenarios. 

However, not all the observations concerning effective visualization designs done in the 2D spatial case can be carried over into the 3D environment~\cite{mota_comparison_2022}.
Therefore, there exists a large number of opportunities to assess the current approaches and set the foundations that will guide the visual design of 3D urban visual analytics systems. In particular, we would like to highlight the great number of opportunities for evaluations concerning the use of visual metaphors and guided explorations described above. 
Furthermore, another important aspect that should be further explored is the interplay between occlusion handling techniques and data analytics.
In fact, strategies like ghosting and deformation are good examples of techniques that fight occlusion by changing visual elements (e.g., shape, position, or transparency) of the physical elements of the scene.
However, these visual elements are often used to depict thematic data (e.g., color, area). 
There is a need to understand what is the impact of the use of such techniques on the overall analytics of thematic data and in which situations each technique is most appropriate.
Finally, many applications, especially the ones focusing on immersive experiences, try to achieve this via rendering realism.
However, similar to the discussion above, the presence of rendering realism (shading, shadows) might interfere with the perception of thematic information.
Again, studying the user's perception in such cases is an interesting direction for future research.

\subsection{Open 3D urban visual analytics}

As we reviewed and categorized papers, our immediate finding was that, while urban visualization in general has made significant advancements in previous years, there is still a notable gap between visualization contributions and domain requirements.
The status quo is one where visualization contributions using 3D urban data are, more often than not, systems that are rarely used beyond the scope of the collaboration that motivated the project in the first place.
In the absence of easy-to-use tools designed with visualization best practices in mind, domain experts resort to using an amalgamation of general tools and libraries, from computational notebooks (e.g., Jupyter Notebooks) to comprehensive GIS tools (e.g., ArcGIS)~\cite{yap_free_2022}.
While this approach might be sufficient to tackle well-defined urban problems in selected regions, the unprecedented challenges facing cities today call for tools and systems that lower the barrier to performing analyses that are reproducible, replicable, and extensible to enable transdisciplinary and convergent research~\cite{ramaswami_sustainable_2018,lobo_convergence_2021}.
Moreover, GIS tools do not consider recent design and evaluation studies and do not provide ways to integrate novel techniques found in visualization studies, hence primarily relying on standard visualization encodings.
If done right, visualization and visual analytics can play a key role in creating the needed collaborative environments that can be translated to different regions and use cases.
To achieve this, there is a need for (1) community building activities that bring together urban experts and visualization researchers, and (2) the adoption of open-source practices by visualization researchers.

To bring together different communities, we believe that the roadmap outlined by Lan et al.~\cite{lan_visualization_2021} for astrophysics might be appropriate for the urban domain. This path involves joint workshops between visualization researchers and urban experts to be held at conferences in the respective areas, such as IEEE VIS and SimAUD.
Examples that can serve as a guideline include ASSETS' UrbanAccess~\cite{froehlich_future_2022}, and VIS' CityVis~\cite{goodwin_unravelling_2021} and EnergyVis workshops.
Another potential activity is the creation of visualization challenges using 3D urban data created by experts.

Regarding the adoption of open-source practices, in our survey, none of the visualization contributions were made publicly available, let alone built a community of contributors. In a recent survey by Yap et al.~\cite{yap_free_2022}, out of 70 open-source tools for urban analytics, only one came from the visualization community~\cite{miranda_urban_2017}, and it did not take into account 3D urban data.
Adjacent urban topics can provide inspiring examples to the visualization community on how to build and sustain open urban tools. Specifically, OSMnx is a widely used Python-based tool for street network analysis~\cite{boeing_right_2020}. The visualization community should consider that it is not enough to build tools, but that the scientific community should strive for public \emph{reusable} tools and workflows.
3D urban visual analytics, much like urban analytics in general, is inherently complex, as the data is oftentimes large, interactions are difficult (especially in 3D), and studies are hard to translate to other regions. While reusability does not seem to be at the forefront of initiatives in the visualization community, well-engineered open efforts, accompanied by outreach efforts, are poised to be embraced by urban experts, given the variability of use cases, data, and tasks. Providing the necessary code base to facilitate urban projects in this domain can lead to not only important domain contributions but also to efforts that provide a more grounded understanding of how experts analyze and visualize 3D urban data.

\subsection{Representations and visual modalities}

\highlight{When it comes to data visualization, the prevalence of 3D representations across examined use cases comes as little surprise. 3D representations provide better contextualization and spatial understanding compared to 2D representations. In particular, such representations allow users to better visualize how complex urban data and the related information interacts with the 3D landscape, especially in relation to the terrain, buildings or other features present in the environment.}
\highlight{Only a handful of works utilize immersive display modalities, but almost all of them make use of a pedestrian view.  Immersive environments provide a 3D spatial context that can closely resemble real-world interaction. This allows users to perceive and understand urban space more naturally, which can be beneficial for planning scenarios. Another path is utilizing immersive environments to visualize and analyze complex spatial phenomena, as for instance, to make sense of micro climate and wind simulations and putting thematic data into spatial context. However, such immersive 3D urban visual analytics solutions require carefully designed interaction and navigation metaphors as opposed to the pedestrian view.}
\highlight{A promising opportunity to bring thematic data or planning situations directly into the actual urban space is by utilizing augmented reality glasses. The XR community has made essential progress in the last decade, such as providing accurate outdoor tracking or dealing with limited hardware capabilities of edge devices.}

\subsection{Collaborative urban analytics platforms}

The process of enabling collaborative urban planning and assessment is pivotal for facilitating effective decision- and policy-making, while also encouraging active community involvement in the urban development processes. However, a limited number of existing urban analytics systems provide a platform for multiple stakeholders to collectively visualize, interact with, and contribute to proposed urban developments and policies. Even among the existing systems that do facilitate such collaboration, the emphasis is often placed on tailoring visualization and interaction features primarily for domain experts and practitioners, with less attention being given to incorporating the requirements of the broader public. However, there is a growing need to integrate these systems into a more inclusive context that caters to the diverse technological skills and information needs of diverse individuals. This will help bridge the gap between expert-driven functionalities and the broader public's participation and will lead to the democratization of urban exploration and planning.

\section{Conclusions}
\label{sec:conclusion}

In this survey, we have presented a detailed review of publications in the area of 3D urban data and visual analytics.
To ensure a diverse corpus of papers, we have engaged in discussions with domain experts, also co-authors of the survey to select appropriate venues in their respective fields of study.
Our selection process was then based on an extensive review of almost 20 venues, including visualization and cross-cutting ones. 
From an initial set of 669 papers, using a voting process where each paper was reviewed by two authors of the survey, we created a corpus of 175 papers covering a period of fifteen years.
To highlight common practices and research opportunities, we framed the survey around three main questions: \emph{Why}  is 3D urban data being analyzed, \emph{What} data is being analyzed, and \emph{How} it is being analyzed and visualized. For each one of these questions, we have derived a set of tags covering use cases, analysis actions and targets, physical and thematic aspects of the data, and visualization and interaction.
The most immediate observation is the growth in the number of publications on the topic; almost half of the surveyed works were published in the last four years. More than that, they cover a wide breadth of visualization contributions, including systems, techniques, design studies, and evaluations.

While previous works have proposed design spaces for immersive urban analytics~\cite{chen_exploring_2017}, primarily focusing on VR, the same cannot be said about general 3D urban data visual analytics. Contributions to the topic are then mostly driven not by the need to explore a design space but by one-off collaborations between visualization researchers and domain experts. This hampers not only the evaluation of different visualization contributions but also the translation of research outcomes to actual workflows across different use cases. 
On top of that, there is a lack of toolkits specifically designed with 3D urban environments in mind. Such toolkits would also facilitate analyses across multiple scales. By considering urban studies, we have found that domain experts oftentimes need to use multiple computational frameworks and libraries to work with data and perform analyses at different scales, increasing data interoperability problems and hampering the analysis workflow.
Moreover, while commercial frameworks, like ArcGIS, offer incredible capabilities for different use cases, they are closed systems, making it hard for visualization researchers to translate their contributions to scientific and engineering domains and workflows.
New visualization contributions to the topic are then required to ``reinvent the wheel'' every time a new problem or use case is tackled -- an incredibly wasteful endeavor given the data and engineering challenges that need to be solved to build a reliable tool.
In this survey, we hope to have established, through a common categorization of papers and detailed discussion, a step towards driving research in visual analytics of 3D urban data. 
Specifically, we hope that the map of visualization and domain contributions can be used by researchers, domain experts, and practitioners as a guideline for new research opportunities.

\vspace{-0.5cm}
\section*{Acknowledgments}
We would like to thank the reviewers for their constructive comments and feedback. This study was supported by the Discovery Partners Institute, National Science Foundation (\#2320261, \#2330565), CNPq (316963/2021-6, 311425/2023-2), and FAPERJ (E-26/202.915/2019, E-26/211.134/2019).
VRVis is funded by BMK, BMDW, Styria, SFG (Steirische Wirtschaftsf\"{o}rderungsgesellschaft m.b.H. SFG), and Vienna Business Agency in the scope of COMET—Competence Centers for Excellent Technologies (879730, 904918), managed by FFG (\"{O}sterreichische Forschungsf\"{o}rderungsgesellschaft).
This research is part of the project Multi-scale Digital Twins for the Urban Environment: From Heartbeats to Cities, which is supported by the Singapore Ministry of Education Academic Research Fund Tier 1. % (Filip)
Claudio Silva is partially supported by NASA, the National Science Foundation, DARPA, and a gift from Adobe. Any opinions, findings, and conclusions or recommendations expressed in this material are those of the authors and do not necessarily reflect the views of Adobe, DARPA, NSF, or NASA.
Copyright for the reprinted images resides with the authors or publishers; images are reprinted with their permission.

% \begin{figure*}[h]
% \begin{center}
% \includesvg[inkscapelatex=false, width=1.00\linewidth]{figs/temp.vega.svg}
% \end{center}
% \end{figure*}

% bibtex
\bibliographystyle{eg-alpha}
\bibliography{urban-3D-ref}

\end{document}